\documentclass[sigconf]{acmart}

\usepackage{subfig}
\usepackage{booktabs}
\usepackage{multirow}
\usepackage{amsmath}
\usepackage{amsfonts}
\usepackage{algorithm}
\usepackage{algpseudocode}
\usepackage{lipsum}  
\usepackage{bm}
\usepackage{soul}
\usepackage{xcolor}
\usepackage[normalem]{ulem}
\usepackage{xifthen}
\usepackage[switch]{lineno}
\newcommand{\modname}{\textsc{Minicorn}}
\newcommand{\longname}{MInimalist Non-reinforced Interactive COnversational Recommender Network}
\useunder{\uline}{\ul}{}

\makeatletter
\newcommand\footnoteref[1]{\protected@xdef\@thefnmark{\ref{#1}}\@footnotemark}
\makeatother

\AtBeginDocument{%
  \providecommand\BibTeX{{%
    \normalfont B\kern-0.5em{\scshape i\kern-0.25em b}\kern-0.8em\TeX}}}
    
\usepackage{lipsum}

\AtBeginDocument{%
  \providecommand\BibTeX{{%
    \normalfont B\kern-0.5em{\scshape i\kern-0.25em b}\kern-0.8em\TeX}}}

\setcopyright{acmcopyright}
\copyrightyear{2018}
\acmYear{2018}
\acmDOI{XXXXXXX.XXXXXXX}

\acmConference[Conference acronym 'XX]{Make sure to enter the correct
  conference title from your rights confirmation emai}{June 03--05,
  2018}{Woodstock, NY}
\acmPrice{15.00}
\acmISBN{978-1-4503-XXXX-X/18/06}

\begin{document}

\title[Minimalist and High-performance Conversational Recommendation]{Minimalist and High-performance Conversational Recommendation with Uncertainty Estimation for User Preference}


\author{Yinan Zhang}
\email{yinan002@e.ntu.edu.sg}
\affiliation{%
  \institution{Nanyang Technological University}
  \country{Singapore}
}

\author{Boyang Li}
\email{boyang.li@ntu.edu.sg}
\affiliation{%
  \institution{Nanyang Technological University}
  \country{Singapore}}

\author{Yong Liu}
\email{stephenliu@ntu.edu.sg}
\affiliation{%
  \institution{LILY Research Centre}
  \institution{Alibaba-NTU Singapore JRI}
  \country{Singapore}}

\author{You Yuan}
\email{youyuan.yy@alibaba-inc.com}
\affiliation{%
  \institution{Alibaba Group}
  \country{China}}

\author{Chunyan Miao}
\email{ascymiao@ntu.edu.sg}
\affiliation{%
  \institution{LILY Research Centre}
  \institution{Nanyang Technological University}
  \institution{Alibaba-NTU Singapore JRI}
  \country{Singapore}
 }
\renewcommand{\shortauthors}{Yinan Zhang and Boyang Li, et al.}

\begin{abstract}
  Conversational recommendation system (CRS) is emerging as a user-friendly way to capture users' dynamic preferences over candidate items and attributes.
  Multi-shot CRS is designed to make recommendations multiple times until the user either accepts the recommendation or leaves at the end of their patience.
  Existing works are trained with reinforcement learning (RL), which may suffer from unstable learning and prohibitively high demands for computing. In this work, we propose a simple and efficient CRS, \longname{} (\modname{}).
  \modname{} models the epistemic uncertainty of the estimated user preference and queries the user for the attribute with the highest uncertainty. The system employs a simple network architecture and makes the query-vs-recommendation decision using a single rule. Somewhat surprisingly, this minimalist approach outperforms state-of-the-art RL methods on three real-world datasets by large margins. We hope that \modname{} will serve as a valuable baseline for future research. 
\end{abstract}

\ccsdesc[500]{Information systems~Recommender systems}

\keywords{conversational recommendation, uncertainty estimation, recommender system}


\maketitle

\section{Introduction}
\begin{figure}
    \centering
    {\includegraphics[width=\columnwidth]{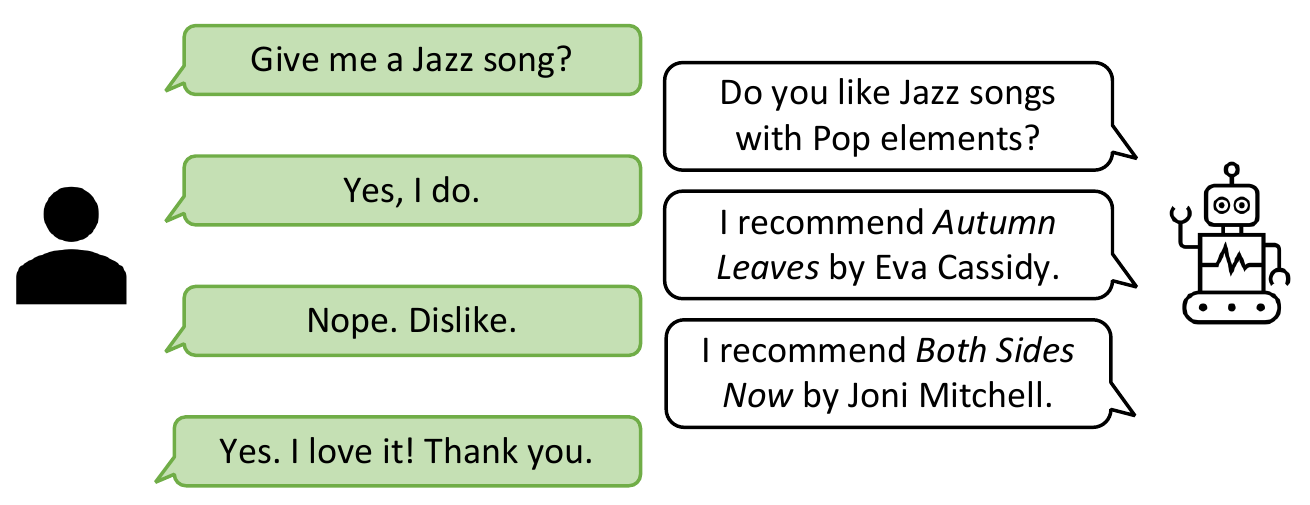}}
    \caption{A conceptual illustration of multi-shot conversational recommendation.}
    \label{fig:workflow}
\end{figure}

With the wide availability of intelligent personal assistants such as Apple Siri and Amazon Alexa, conversational recommendation system (CRS) is emerging as a new paradigm for recommendation systems \cite{CRSAL,yongeng2018}. 
Traditional recommendation systems derive recommendations directly from historic interactions with the user, including explicit \cite{ExplicitRec1,ExplicitRec2,ExplicitRec3,ExplicitRec4} and implicit feedback \cite{ImplicitRec1,ImplicitRec2,ImplicitRec3,ImplicitRec4}. However, historic information may become obsolete when user preference changes since the last interaction with the system. In cold-start situations, little historic information is available for recommendation decisions. In addition, privacy concerns may prevent the collection and storage of excessive interaction data. In these cases, it may be advantageous to directly ask users for their preferences in the form of conversations.  Finally, CRS affords a natural, speech-based form of interactivity, which is convenient when the user is driving or running, and necessary when the user is visually impaired.

\sloppypar{The conversational recommendation (CR) problem consists of two major sub-problems, both of which pose serious challenges and have received substantial research attention. The first one is to understand potentially ambiguous user responses in spoken language and extract discrete preferences (e.g., like/dislike) \cite{CRM,chen2019towards,Raymond18}. The second sub-problem is to devise a conversational strategy, such as one that determines when to collect more information and when to make a recommendation \cite{jie20towards,EAR,SCPR,KBQG}. To simplify the problem, in this paper, we tackle the second problem and assume the user responses have been disambiguated; we aim to develop a conversation strategy maximizes recommendation accuracy in the least rounds of conversations.}

Following \cite{EAR,SCPR}, we investigate \emph{dynamic multi-shot CRS}. 
With Figure \ref{fig:workflow}, we illustrate the intended use case of multi-shot CRS \cite{EAR,SCPR}. The user initiates a recommendation request and supplies a desired item attribute (e.g., ``Jazz''). At each round of conversation, the CRS either asks the user for preference of an attribute (e.g., ``Pop'') or recommends one or more items. The conversation continues until the user accepts one recommendation or leaves the conversation. 
Each item is associated with a set of attributes, such as the genres of a movie or the language of a song. Users' answers to attribute questions can be used to improves future recommendations and to narrow down the candidate item set. CRS may make recommendations at more than one rounds and recommend multiple items at one round.

Reinforcement learning (RL) has been the method of choice for multi-shot CRS \cite{EAR,SCPR,KBQG} for its ability to optimize sequences of discrete actions without direct supervision. RL may seem necessary as recommendation datasets typically do not have ground-truth annotations for the attribute questions or question-versus-recommendation decision. However, RL also has high computational requirements \cite{RL_requirement}, high sample complexity \cite{gu2017qprop,nachum2018trustpcl}, and sensitivity to hyperparameters \cite{henderson2019deep,islam2017reproducibility,haarnoja2018soft}, which could complicate training. 

With this paper, we demonstrate a simple, effective, and practical CRS that does not rely on reinforcement learning. The proposed system, named \longname{} (\modname{}), outperforms existing state-of-the-art solutions \cite{EAR,SCPR} that are end-to-end optimized with RL. 

The efficiency of user preference acquisition and estimation is crucial in CRS. In order to minimize the number of conversation turns, each user query should be as informative as possible. To this end, we design the following system components: First, we quantify the uncertainty in estimated user preference over attributes and select the attribute with the highest uncertainty for questioning. This is in contrast to existing works that ask questions most likely to receive affirmative answers \cite{QandA,CRM,EAR,KBQG} or questions that eliminate as many candidate items as possible \cite{SCPR}. Second, we model the correlation among attributes. This allows \modname{} to deduce, from known user responses, preference over attributes that have not been queried yet. Experiments demonstrate the crucial roles played by these two system components.

With this paper, we make the following contributions:
\begin{itemize}
    \item We present a simple method with high performance. \modname{} employs multi-layer perceptrons and convolution with few bells and whistles like reinforcement learning or complex attention mechanisms. \modname{} makes the query-vs-recommendation decision using a single rule. It is thus perhaps surprising that \modname{} outperforms the previous state of the art by up to 67.40\% and 13.39\% in recommendation success rate and conversation turns. The simplicity of \modname{} makes it immensely practical for industrial application and a valuable baseline for future research. 
    \item To our knowledge, \modname{} is the first work to model the uncertainty of predicted user preference in multi-shot CRS. Since user preference is a major source of epistemic uncertainty in CRS, quantification of the uncertainty facilitates the creation of effective user queries. 
    \item We present a straightforward method for modeling the correlation among item attributes, which boosts the information content acquired from each user response. 

\end{itemize}

\section{Related Work}
\subsection{Conversational Recommendation Systems}

With broad strokes, we may categorize existing conversational recommendation systems (CRSs) \cite{Wenqiang2021,Venkata2021} into those that handle natural language input and those that focus solely on recommendation. In the former category, \cite{CRM,Christakopoulou2016,liang2021learning,heming2021,Wenchang2020,conv_tianshu2022,conv_Papenmeier2022} parse natural language into a number of predefined slots such as the location and price range of a restaurant. 
\cite{Raymond18,chen2019towards,Yan2017,wang2021finetuning,Krisztian2021,xiao2021endtoend,yu2021} develop end-to-end conversational shopping assistants that perform a variety of tasks including recommendation.

Among recommendation-oriented CRS, we make the distinction between those with fixed conversation strategies and those with dynamic strategies. In the first type of fixed strategies, the CRS asks a fixed number of questions and makes a recommendation at the last conversational turn \cite{CRM,yongeng2018,Christakopoulou2016}. In the second type, the CRS always asks a question and makes recommendations at every turn \cite{QandA,chen2019towards,Montazeralghaem2021}. In both types, the CRS does not make conversation-dependent choices between questioning and recommendation. 

In comparison, the dynamic strategy setting requires the 
CRS to choose between raising a question and recommending items. Typically, the CRS makes multiple recommendations (i.e., \emph{multi-shot} CRS) until user acceptance or a predefined maximum number of turns has elapsed. This setup gives CRS more freedom and arguably represents more realistic modeling of conversations. 

EAR \cite{EAR} establishes a unified dynamic multi-shot CRS framework with three sub-problems: 1) what attribute to query, 2) when to switch from user query to recommendation, and 3) how to adapt to user feedback.
EAR-FPAN \cite{FPAN} improves the recommendation module of EAR by modeling the relation between attribute-level and item-level feedback signals.
SCPR \cite{SCPR} and others \cite{KBQG,yang2021} employ the user-item-attribute graph to model user preference. \cite{zhao2021knowledge,conv_yuanhang2022,conv_zhao2022} use the knowledge graph derived from external data sources.
As the CRS performs discrete actions, existing multi-shot CRSs \cite{EAR,FPAN,SCPR,KBQG,zhao2021knowledge,yang2021} are all trained with reinforcement learning. 

Regarding the type of user queries, \cite{Yaxiong2021,Xiaoxiao2018} require users to comment on the recommended items. \cite{EAR,SCPR,zhao2021knowledge} adopt a yes/no question for each query about a single attribute.  \cite{KBQG,yang2021,conv_rec_yiming2022} allows open-ended questions about an attribute type such as ``which movie genre would you like?'' Though open-ended questions are informative, the user response may not readily fall into predefined attributes known to the system. 
In this work, we adopt yes/no questions and compare against methods \cite{EAR,SCPR} that use the same question type and do not use external information sources. 

Existing systems also adopt different questioning strategies. 
\cite{SCPR,yang2021} generate queries that cause maximum reduction of the probability mass over recommendation candidates. \cite{EAR,FPAN,KBQG} choose the attribute that the user is most likely to favor.

An area related to CRS is conversational search systems \cite{conv_survey,conv_xing2022}. Research work in this area tend to focus on resolving ambiguity in natural language \cite{conv_Frummet2022,conv_classification_Lili20,amb_Salle2022}, on open-domain large-scale document retrieval \cite{doc_Ivan2022,doc_Rafael2022,conv_Ivan2022,doc_Kim2022}, and often do not to utilize historic user-item interactions \cite{conv_liao2021,conv_zhenduo2021,conv_Kiesel2021}. Conversational recommendation systems consider both users' current request and historical interactions.

\subsection{Uncertainty Estimation}

Due to their sheer sizes, deep neural networks pose special challenges to uncertainty quantification, as traditional methods incur prohibitive computational cost \cite{osawa2019practical,rajaratnam2015mcmcbased}.
Deep Ensemble \cite{Lakshminarayanan2017} utilizes inherent stochasticity in NN training to create multiple networks. Techniques such as Monte Carlo (MC) Dropout \cite{Srivastava2014,dropout}, Vertical Voting \cite{xie2013horizontal}, TreeNet \cite{lee2015m}, Batch Ensemble \cite{wen2020batchensemble}, Multi-input Multi-output Ensemble \cite{havasi2021training}, and Orthogonal Dropout \cite{zhang2021ex} can be seen as carving out diverse subnetworks from one large neural network. Multiple networks may also be created from a single training session \cite{huang2017snapshot}. 
From the outputs of diverse networks or subnetworks, uncertainty estimates can be derived. 

A rich body of work extends the Bayesian approach, which provides out-of-the-box uncertainty estimates, to deep networks \cite{Kendall2017,malinin2018predictive,lee2018deep,mandt2018stochastic}. 
Approaches could be assorted into variational Bayes \cite{louizos2017,ashukha2020pitfalls,osawa2019practical} and stochastic gradient MC sampling \cite{Welling2011,li2016preconditioned,heek2020bayesian}. To this day, accurate and efficient Bayesian inference remains an open research problem \cite{Yao2019,wenzel2020good}. 

In this work, we focus uncertainty estimation on the prediction of user preference over item attributes, which is a major source of epistemic uncertainty as the system attempts to actively acquire user preferences. Though we mainly utilize MC Dropout \cite{dropout}, our method is compatible with other static ensemble-based techniques.

\section{Problem Definition}
We first present essential notations. We use 
$u\in \mathcal{U}$ to denote the user and $v \in \mathcal{V}$ to denote the item. 
Each item $v$ is associated with a set of attributes $\mathcal{P}_v \subset \mathcal{P}$ ($\mathcal{P}_v \neq \emptyset$), such as "Comedy" or "Thriller" for a movie. $\mathcal{U}, \mathcal{V}$, and $\mathcal{P}$ represent the universes of users, items, and attributes, respectively. 
We use binary vector $\mathbf{b}(v) \in \{0,1\}^P$ to represent attributes for item $v$, where $P = |\mathcal{P}|$ is the total number of attributes.
If $p\in \mathcal{P}_v$, then the $p^{\text{th}}$ element in $\mathbf{b}(v)$ is assigned to 1. Otherwise, it is set to 0. $\mathcal{V}[p]$ denotes the set of items that has attribute $p$. That is, $\mathcal{V}[p] = \{v | \mathbf{b}(v)_p = 1 \}$. 

We follow the dynamic multi-shot CRS scenario as studied in \cite{EAR,SCPR,KBQG}. The CRS may ask questions or recommend items to users at every conversational turn until the users accept the recommendation or the conversation reaches a predefined maximum number of turns. The goal is to recommend the desired item within as few conversational turns as possible.

The multi-shot conversational recommendation is initiated by users with the assumption that the user $u$ is desired for a specific item $v$ and the user responses are unambiguous. In the first conversational turn, the user $u$ provides a preferred attribute $p_1 \in \mathcal{P}_v$ based on the desired item $v$.
Afterward, the CRS leads the conversation by asking questions or recommending items at the turns $t=2,3,\cdots,T$, where $T\leq T_{max}$ denotes the last turn of the conversation and $T_{max}$ is the longest conversation rounds that the user can endure.
The question is designed to clarify user preference on one specific attribute. As illustrated in Figure \ref{fig:workflow}, the query ``Do you like Jazz songs with Pop elements?'' is pushed to clarify whether the user likes the attribute ``Pop''.
Each round of recommendation $\mathcal{R}_t$ contains $K$ items (i.e., $|\mathcal{R}_t|=K$).
The conversation will be terminated if the user accepts the recommendation ($v\in \mathcal{R}_t$) or runs out of patience ($t\geq T_{max}$).

\section{The \modname{} Method}
\begin{figure*}[t!]
  \centering
  \includegraphics[width=2\columnwidth]{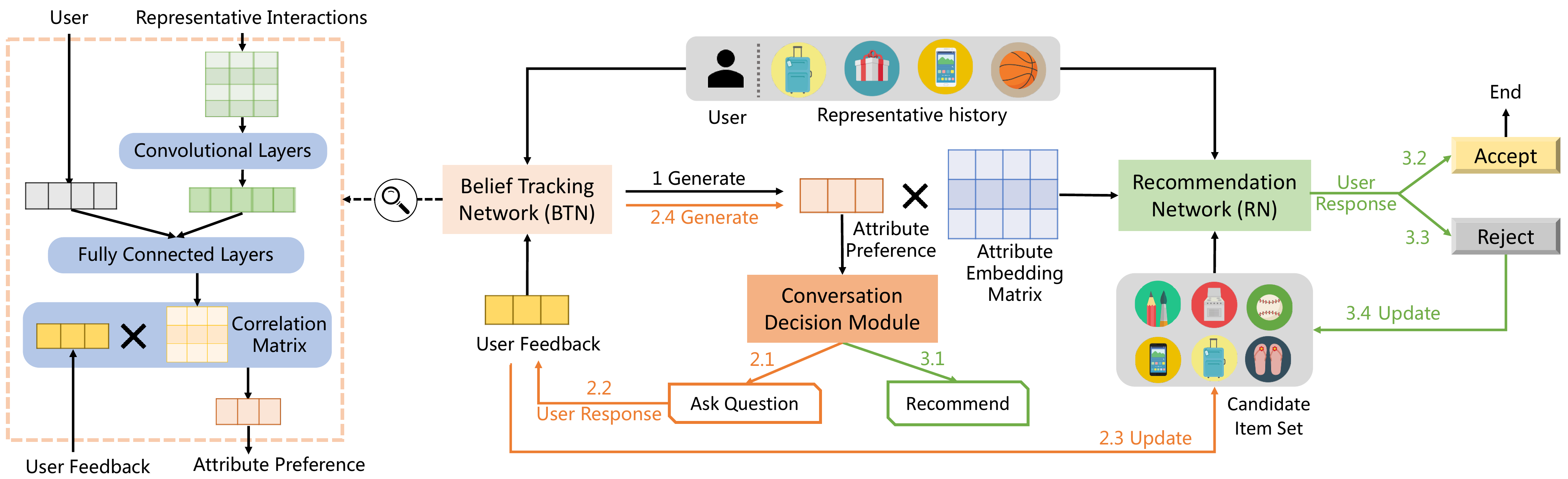} 
  \caption{The architecture of \modname{}. Belief Tracking Network predicts the user's preference over attributes. Recommendation Network recommends items based on current preference estimates, user, and historic interactions. Conversation Decision Module chooses to either ask an attribute question or recommend items. The dashed box shows the structure of BTN. }
  \label{model}
\end{figure*}

Figure \ref{model} shows the overall system architecture of \modname{}, including the Belief Tracking Network, the Recommendation Network, and the Conversation Decision Module. The workflow steps are indicated by the numbers attached to the arrows. 

We start every turn of the conversation with the estimated attribute preference, $\mathbf{q}_{t-1}$, from the last turn. $\mathbf{q}_{t-1}$ is a $P$-dimensional vector with values between 0 and 1. A value of 1 indicates the network is confident that the corresponding attribute is desirable. 0 indicates that the network is confident that the attribute is undesirable. A value between 0 and 1 indicates some degree of ambivalence. 

With $\mathbf{q}_{t-1}$ as the current belief over attributes, the Conversation Decision Module may choose either to query the user about preference on an attribute (Step 2.1) or to make a recommendation (Step 3.1). After the user responds to the query, \modname{} updates the user feedback vector $\mathbf{a}_t$ (Step 2.2) and updates the candidate item set $\mathcal{V}_t$ (Step 2.3). The conversation then moves to the next turn and CRS estimates user preferences, $\mathbf{q}_{t}$, based on the latest updated user feedback $\mathbf{a}_t$ (Step 2.4).

To make a recommendation, the Recommendation Network $\mathcal{S}$ scores all candidate items and recommends $K$ best ranked items.
If the user accepts the recommendation (Step 3.2), the task succeeds and no further conversation is necessary. Otherwise, if the user rejects the recommendation (Step 3.3), \modname{} will remove the rejected items from the candidates (Step 3.4). We then start the next turn of the conversation. The conversation always ends before or at turn $T_{\text{max}}$, which represents the maximum patience of the user.

\subsection{Conversation Decision Module}

By choosing between the query action and the recommendation action, the Conversation Decision Module plays a crucial role in CRS. Excessive information collection may exhaust user's patience, whereas premature recommendation may lead to poor accuracy as the system lacks understanding of the user. The overall objective of CRS is to make the most accurate recommendations in the least number of conversation turns.

In contrast to policies learned using reinforcement learning (RL), \modname{} applies a simple rule for the conversation decision: If we are not sufficiently confident about the preference of at least one attribute and we have not reached the last turn and the candidate item size is larger than the number of items in a single recommendation, ask a question. Otherwise, make recommendations. As we show in the experiments, this simple rule performs surprisingly better than several RL methods. 

With the preference belief vector $\mathbf{q}_{t-1}$, the confidence $C(p)$ over the preference of attribute $p$ is computed as distance from the midpoint of the range (0, 1): 
\begin{equation}
    C(p) = |q_{t-1, p}-0.5|,
\end{equation}
where $q_{t-1, p}$ refers to the $p^{\text{th}}$ element in the vector $\mathbf{q}_{t-1}$. The decision rule is controlled by a hyperparameter $\alpha$. If $C(p) \le \alpha$ for any attribute $p$ , $t<T_{max}$, and $|\mathcal{V}_{t-1}| > K$, \modname{} chooses the query action. Otherwise, it makes recommendations. 

In the following, we will explain how \modname{} creates the query and will leave details of the recommendation to Section \ref{sec:rec}. To find an attribute to ask the user about, \modname{} chooses the attribute with the most uncertainty, which is supposed to provide the most information. This differs from existing work that selects the attribute that the user will most probably find desirable \cite{QandA,CRM,EAR,KBQG} or the attribute leading to the maximum candidate items reduction \cite{SCPR}.

In the first measure of uncertainty, Monte Carlo (MC) Dropout \cite{dropout}, we compute the degree of change in the belief vector $\mathbf{q}_{t-1}$ when some intermediate representations in the Belief Tracking Network (BTN) are erased. We perform $N$ forward passes with BTN, each time with a different dropout mask, resulting in $N$ different beliefs $\mathbf{q}_{t-1}^{(1)}, \mathbf{q}_{t-1}^{(2)},\ldots, \mathbf{q}_{t-1}^{(N)}$. After that, we calculate the element-wise variance, $\sigma_p$, as
\begin{equation}
\label{mcdropout}
    \sigma_p = \frac{1}{N} \sum_{n=1}^N \left( q_{t-1, p}^{(n)} - \mu _p\right)^2, \, \, \mu_p=\frac{1}{N}\sum_{n=1}^N q_{t-1, p}^{(n)}.
\end{equation}
The variance vector $\bm{\sigma} = [\sigma_1, \sigma_2, \cdots, \sigma_P]$ is normalized such that all $\sigma_p$ falls in the range $[0, 1]$. 

We complement the first uncertainty measure with a second measure, which we call Midpoint Proximity. It is computed as the proximity to the midpoint of the belief range, 0.5. 
\begin{equation}
\label{midpoint}
    \mathbf{r} = \text{Norm} \left(1 - 2 |\mathbf{q}_{t-1}-0.5 \cdot \bf{1}|\right),
\end{equation}
where $\bf{1} \in \mathbb{R}^P$ is an all-one vector. We again apply the $\text{Norm}(\cdot)$ operation, which normalizes components of the vector $\mathbf{u}$ to the range $[0, 1]$. Finally, we combine the two measures using element-wise harmonic mean, $2 r_p \sigma_p / (r_p + \sigma_p)$. \modname{} queries the user about the preference on the attribute with the highest uncertainty. 

After receiving the user response, we update the user feedback vector $\bm{a}_t \in \{0, 0.5, 1\}^P$. The vector is initialized as $0.5 \cdot \bm{1}$, indicating unknown preferences. When a user gives a favorable response to an attribute query, the corresponding element is set to 1. With an unfavorable response, the element is set to 0.

\subsection{Belief Tracking Network}
The Belief Tracking Network (BTN) predicts the user preference vector $\mathbf{q}_t$. Its input includes the user embedding $\mathbf{e}^{\text{user}}_u\in \mathbb{R}^D$, binary attribute matrix $B_u \in \{0,1\}^{|\mathcal{H}_u|\times P}$, and the user feedback vector $\mathbf{a}_{t}$. Here the history set $\mathcal{H}_u$ denotes a set of items the user $u$ has interacted with previously. Note that we may not use the entire history and may select a few representative items. Each row in the matrix $B$ is a binary Bag-of-attributes representation for one item in $\mathcal{H}_u$. In function form, we can write $\mathbf{q}_t = \text{BTN}\left(\mathbf{e}^{\text{user}}_u, B_u, \mathbf{a}_t\right)$.

To minimize the number of queries, we exploit the correlation between attributes to infer real preferences from limited user responses. Such correlation may be highly informative. For example, in the MovieLens dataset \cite{movielens}, out of the 104 movies with the attribute \emph{Animation}, 80\% have the attribute \emph{Children's} and none has the attribute \emph{Crime}.

The main role of BTN is to model such correlations as a symmetric matrix $A\in \mathbb{R}^{P\times P}$. To enforce symmetry, we let  
$A = \frac{1}{2}(\tilde{A} + \tilde{A}^{\top})$, where matrix $\tilde{A}$ is a function of the user embedding $\mathbf{e}^{\text{user}}_u$ and the history matrix $B_u$. After that, we set the diagonal of $A$ to 1 as to preserve user feedbacks.
With $A$, the preference belief vector $\mathbf{q}_t = A \bm{a}_t$. 

Figure \ref{model} shows the architecture of BTN. 
We feed the history attribute matrix $B_u$ into 2D convolutional layers, and concatenate the output with $\mathbf{e}^{\text{user}}_u$. The concatenation is then fed into fully connected layers. The output is reshaped to a $P$-by-$P$ matrix denoted as $\tilde{A}$. 

\subsection{Recommendation Network}
\label{sec:rec}
The Recommendation Network (RN) takes as input the user embedding $\mathbf{e}^{\text{user}}_u$, the embeddings of historic items $H_u \in \mathbb{R}^{|\mathcal{H}_u| \times D}$, the belief embedding $\bm{o}_t \in \mathbb{R}^D$, and the embedding of one candidate item $\mathbf{e}^{\text{item}}_{v} \in \mathbb{R}^D$. Here, for every item $v$, we create a trainable $D$-dimensional embedding $\mathbf{e}^{\text{item}}_{v}$. The rows of the matrix $H_u$ are embeddings of items in the history $\mathcal{H}_u$. 
The network outputs a score $s(u, v)$ for every user-item pair $(u, v)$. In function form, we write $s(u, v) = \text{RN}(\mathbf{e}^{\text{user}}_u, H_u, \mathbf{e}^{item}_v, \bm{o}_t)$. Note that we use lower score to indicate better preference. In each recommendation action, we select the $K$ candidate items with lowest scores for recommendation. 

In order to relate the belief vector $\mathbf{q}_t$ with item embeddings, we create the belief embedding $\bm{o}_t \in \mathbb{R}^D$ using the following procedure. For each attribute $p$, we compute its embedding $\bm{e}^{\text{attr}}_p$ by averaging over items associated with $p$:
\begin{equation}
    \bm{e}^{\text{attr}}_p = \frac{1}{|\mathcal{V}_p|} \sum_{v\in \mathcal{V}_p}\bm{e}^{\text{item}}_v.
\label{attribute_emebdding}
\end{equation}
The belief embedding $\bm{o}_t$ is computed as a weighted sum of attribute embeddings,
\begin{equation}
    \bm{o}_t = \sum_{p=1}^P \, q_{t-1, p} \,  \bm{e}^{\text{attr}}_p.
\label{belief_emebdding_1}
\end{equation}
In matrix form, this can be written as
\begin{equation}
    \bm{o}_t = \mathbf{q}_{t-1}^\top \; E^{\text{attr}} 
\label{belief_matrix}
\end{equation}
where the rows of matrix $E^{\text{attr}} \in \mathbb{R}^{P \times D}$ are attribute embeddings. 

We introduce the detailed architecture of RN.
We append $\mathbf{e}^{\text{user}}_u$ as a new row to $H_u$ and form a matrix of dimension $(|\mathcal{H}_u|+1) \times D$, which is fed into a Residual Network \cite{resblock}. The network stacks several residual blocks, each containing two 2D convolutional layers. 
Then we feed the concatenation of the output with $\mathbf{o}_t$ into fully connected layers to generate $\mathbf{s}_{t}$, which represents the user's current preference.
Both $\mathbf{s}_{t}$ and $\mathbf{e}^{item}_{v}$ are fed into fully connected layers to evaluate the distance between the user's preference and the evaluated item.

\subsection{Model Training}
\label{sec:training}
In the training of CRS, it is necessary to convert conventional recommendation datasets to supervisory signals that CRS can accept. 
Typically, recommendation datasets provide only a list of user-item pairs $(u,v)$, each showing that user $u$ interacted with item $v$. No records exist regarding if the user searched for certain attributes before finding the desired item.
To train CRS on conventional datasets, researchers \cite{EAR,SCPR,KBQG} usually assume that the user $u$ answers affirmatively to all attributes associated with the target item $v$. 

Therefore, we train the output of BTN, $\mathbf{q}_t$, to match the ground-truth attribute vector $\bm{b}(v)$ regardless of the conversational turn $t$. More formally, we adopt the following loss, containing one binary cross-entropy for each attribute,
\begin{equation}
\label{attribute_loss}
    \mathcal{L}_{Attr} = \mathbb{E}_{u, v, t} \biggl[ \sum_{p=1}^P b(v)_p \log q_{t, p} + \left(1 - b(v)_p\right)\log \left(1 - q_{t, p}\right) \biggr].
\end{equation}

In \modname{}, the Recommendation Network (RN) is trained separately from BTN, which leaves the question of what attribute vector to feed to RN during training. As the BTN prediction $\mathbf{q}_t$ differs from the ground-truth $\bm{b}(v)$, directly using $\bm{b}(v)$ as the input to RN will create a training-test distribution shift. On the other hand, feeding $\mathbf{q}_t$ to RN may create instability in training since $\mathbf{q}_t$ changes as BTN is updated during training. To solve this problem, we create a new attribute vector $\bm{b}^\prime(v)$ as input to RN by randomly masking dimensions of $\bm{b}(v)$ with $0.5$ probability. If a dimension is masked, it is set to 0.5. Otherwise, the dimension retains its ground truth value of either 0 or 1. It is worth noting that the predicted attribute $\mathbf{q}_t$ is used during testing. 

Consequently, the RN function used during training is $s(u, v) = \text{RN}(\mathbf{e}^{\text{user}}_u, H_u, \mathbf{e}^{item}_v, \bm{o}^\prime(v))$, where $\bm{o}^\prime(v) = \bm{b}^\prime(v)^\top E^{\text{attr}}$, and $s(u, v)$ is a scalar score for every $(u, v)$ pair. We apply the following loss that pushes the score of the correct pair $(u, v)$ toward 0 and the score of randomly sampled negative pair $(u, v_{\text{neg}})$ above margin $m$,
\begin{equation}
\label{rec_loss}
    \mathcal{L}_{Rec} =
    \mathbb{E}_{u, v, v_{\text{neg}}}\biggl[ s^2(u, v) +  \left(\max( m - s(u, v_{\text{neg}}), 0)\right)^2 \biggr].
\end{equation}

\section{Experiments}
\subsection{Experimental Settings}
In this work, we initialize the user and item embeddings using \textsc{Leporid} \cite{leporid}.
For each user $u$, the representative history $\mathcal{H}_u$ contains 5 items.
The conversational recommendation session will be terminated by the longest conversation round $T_{max}=15$ and the CRS selects $K=10$ items for each recommendation. 

\subsubsection{Datasets}
\begin{table}[t]
\caption{Dataset Statistics.}
\label{exp:datasets}
\resizebox{\columnwidth}{!}{
\begin{tabular}{@{}cccccc@{}}
\toprule
Datasets    & Users & Items & {Interactions} & {Attributes} & {Avg. Items per Attribute} \\ \midrule
\textit{LastFM} & 1,801   & 7,432    & 76,693 & 33 & 918 (12.35\%) \\
\textit{MovieLens} & 5,745    & 3,549    & 789,079 & 18 &333 (9.38\%)\\
\textit{Yelp} & 27,675   & 70,311    & 1,368,606 & 590 & 808 (1.15\%)\\ \bottomrule
\end{tabular}
}
\end{table}

The proposed \modname{} are evaluated on three real-world datasets: \textit{LastFM}\footnote{\url{https://grouplens.org/datasets/hetrec-2011/}}, \textit{MovieLens} \cite{movielens}, and \textit{Yelp}\footnote{\url{https://www.yelp.com/dataset/}}. 
Following \cite{EAR,SCPR}, we use the manually merged 33 coarse-grained attributes in \textit{LastFM} for the ease of modeling. For \textit{MovieLens} and \textit{Yelp}, we utilize the original 18 and 590 attributes. 
On \textit{MovieLens} and \textit{Yelp}, we consider the latest interactions as the representative history.
On \textit{LastFM}, the interaction time is unavailable, so we include in $\mathcal{H}_u$ the songs that the user listened to most often.
We randomly split the three datasets in the ratio of 7:1.5:1.5 for training, validation, and test. 
We report dataset statistics in Table \ref{exp:datasets}.

\subsubsection{User Simulation}
The conversational recommendation requires feedback from the users. 
Following \cite{EAR,SCPR,KBQG}, we assume the user will provide positive answers to all ground-truth attributes associated with target items and negatively to all other attributes. In the first conversational turn, the user initiates interaction with a random ground-truth attribute.

\subsubsection{Baselines}
\label{baseline}
We select the following traditional and state-of-the-art methods for multi-shot CRS. We use the published codes for implementing baseline methods.
\begin{itemize}
    \item \textbf{Greedy}\footnote{\label{ear_github}https://ear-conv-rec.github.io/} \cite{Christakopoulou2016}, which makes recommendations at every turn based on the user profile and history and does not raise any questions about attributes. 
    \item \textbf{CRM}\footnoteref{ear_github} \cite{CRM,EAR}, which is trained using reinforcement learning. We use the multi-shot version of CRM from \cite{EAR}.
    \item \textbf{Max Entropy}\footnoteref{ear_github}, a rule-based model based on CRM \cite{CRM}. It asks questions every turn until one of the three conditions is met: (1) the user preferences over all the attributes are known; (2) the conversation length reaches $T_\text{max}$; (3) remaining candidate items are fewer than $K$.
    Given attribute $p$, the current candidate set $\mathcal{V}_t$, and all items with attribute $p$, $\mathcal{V}[p]$, the method computes entropy as
    \begin{equation}
    \label{eq:maxentropy}
        - Pr(p) \log Pr(p) - (1-Pr(p))\log (1-Pr(p)),
    \end{equation}
    where $Pr(p) = |\mathcal{V}_t \cap \mathcal{V}[p]| / |\mathcal{V}_t|$, or the proportion of candidate items with attribute $p$ in all candidates. The entropy is maximized when exactly half the candidates have attribute $p$. In this case, either the positive feedback or the negative feedback will eliminate half the candidates. Therefore, the attribute with the highest entropy can be understood as the attribute that reduces the candidate set the most in expectation. 
    The baseline uses the same network architecture and training procedure for recommendation as CRM \cite{CRM}.
    \item \textbf{EAR}\footnoteref{ear_github} \cite{EAR}, which is a RL-based method and queries with the attribute that the user is most likely to like.
    \item \textbf{EAR-FPAN}\footnote{https://github.com/xxkkrr/FPAN} \cite{FPAN}, which improves the recommendation component of EAR while keeping the conversational component of EAR unchanged. It utilizes RL to decide whether to raise queries or to make recommendations.
    \item \textbf{SCPR}\footnote{https://cpr-conv-rec.github.io/} \cite{SCPR}, which is a state-of-the-art RL-based method. It always asks the attribute which leads to the maximum reduction in the candidate item set.
\end{itemize}

\subsubsection{Evaluation Metrics}
Following \cite{EAR,SCPR,KBQG,FPAN}, we evaluate the performances of \modname{} using cumulative success rate (SR@T) and average turn (AT). SR@T refers to the accumulative recommendation success rate by turn T. For example, if the user accepts the recommendation at the 5$^{th}$ turn, then we have SR@4=0 and SR@5=1. Higher SR@T represents the better performance. AT is defined as the average number of turns needed for successful recommendation. If the CRS cannot make successful recommendation before the maximum number of turns, we use $T_{max}$ to compute AT. Lower AT indicates better efficiency of CRS.

\subsection{Results and Discussion}
\subsubsection{Recommendation Performance}

\begin{figure*}
    \centering
    \subfloat[\centering\label{fig:LastFM_baseline}]{{\includegraphics[width=2\columnwidth/3]{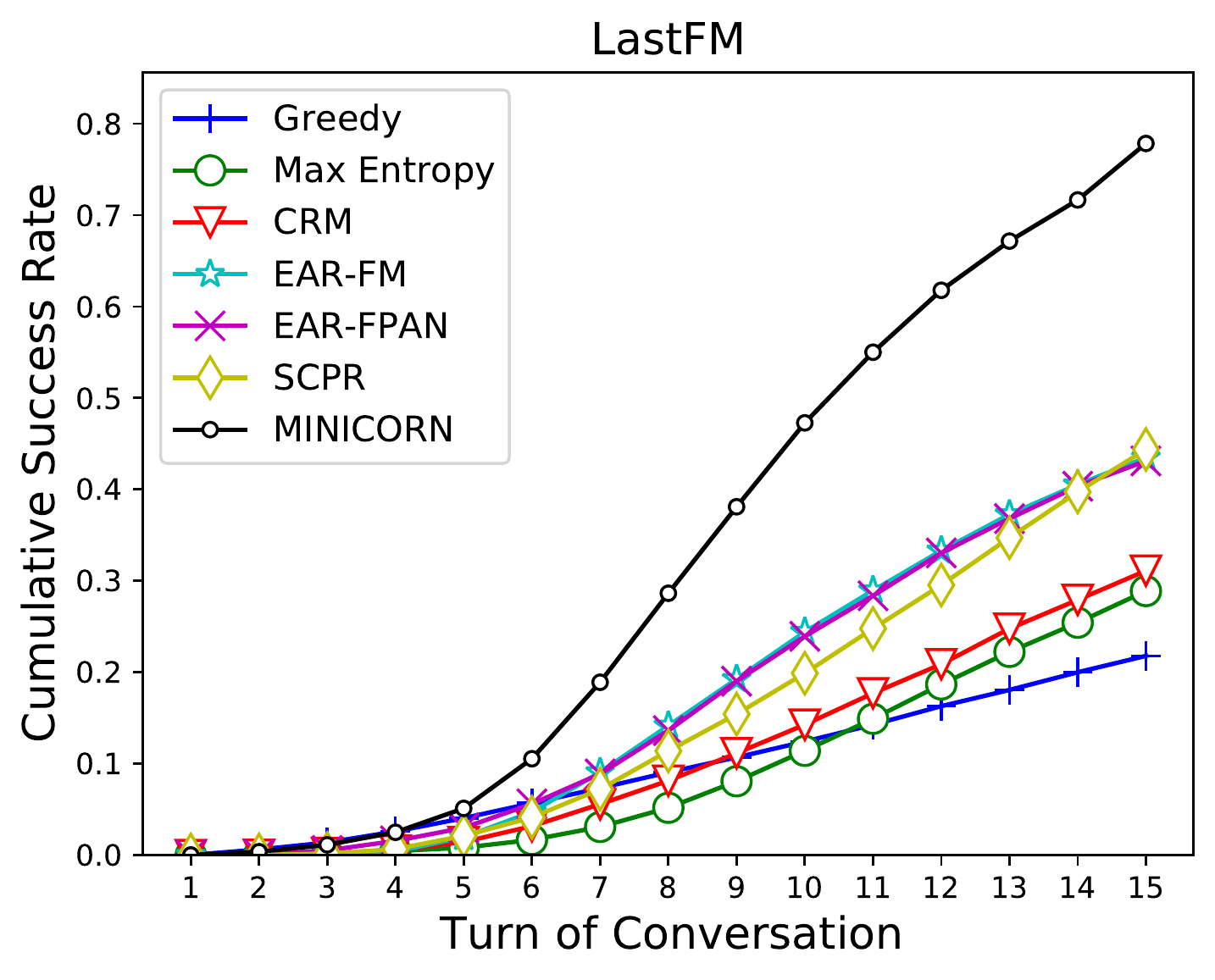}\Description[A line chart]{A line chart comparing different CRS, \modname{} performs the best cumulative success rate from the 5$^{th}$ turn, then achieves almost twice the accuracy of the best baseline. EAR, EAR-FPAN, and SCPR perform comparably.}}} 
    \subfloat[\centering\label{fig:MovieLens_baseline}]{{\includegraphics[width=2\columnwidth/3]{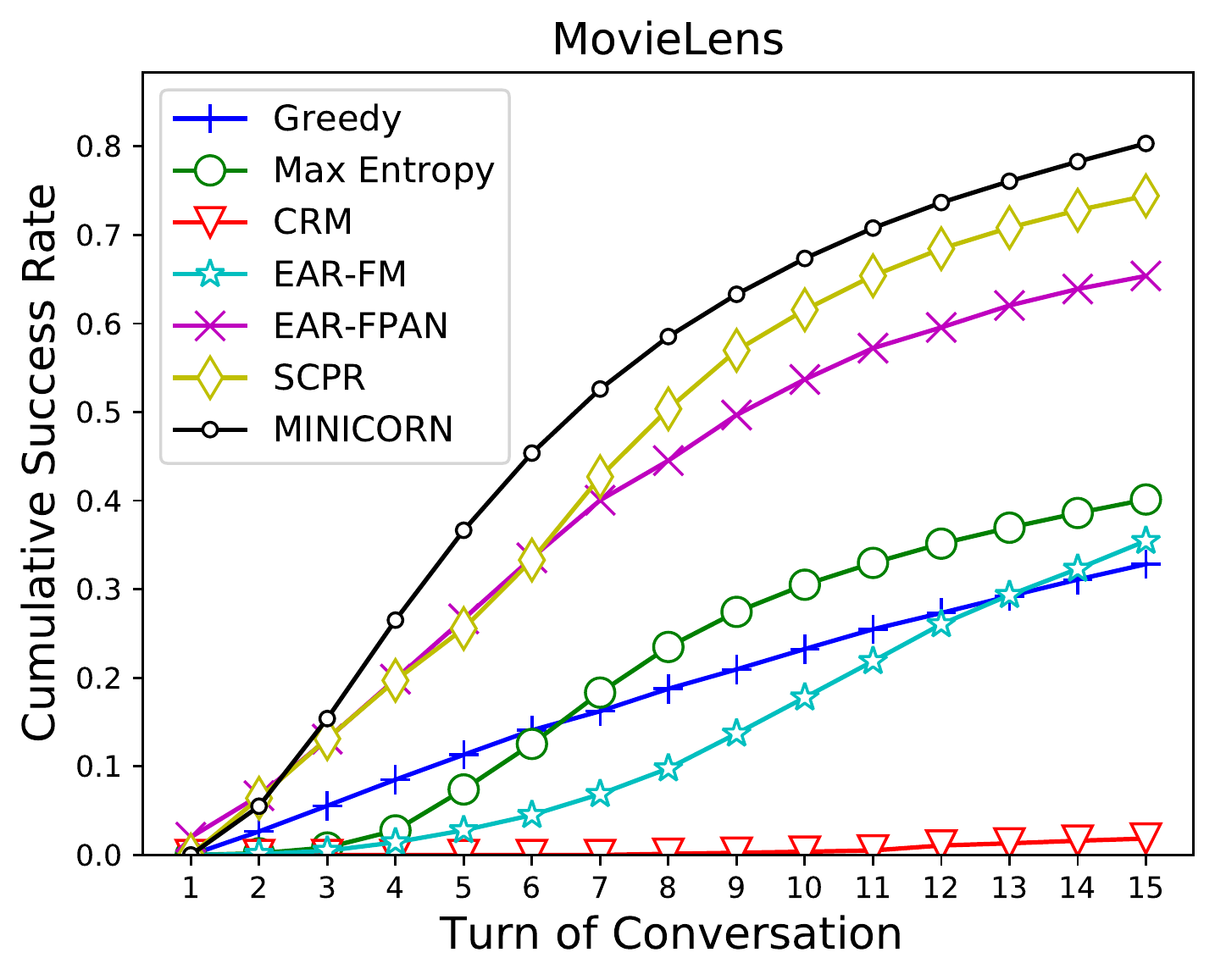} \Description[A line chart]{A line chart comparing different CRS, \modname{} achieves the best cumulative success rate from the 3$^{rd}$ turn and then outperforms the best baselines (SCPR) by about 0.07 of cumulative success rate.}}}
    \subfloat[\centering\label{fig:Yelp_baseline}]{{\includegraphics[width=2\columnwidth/3]{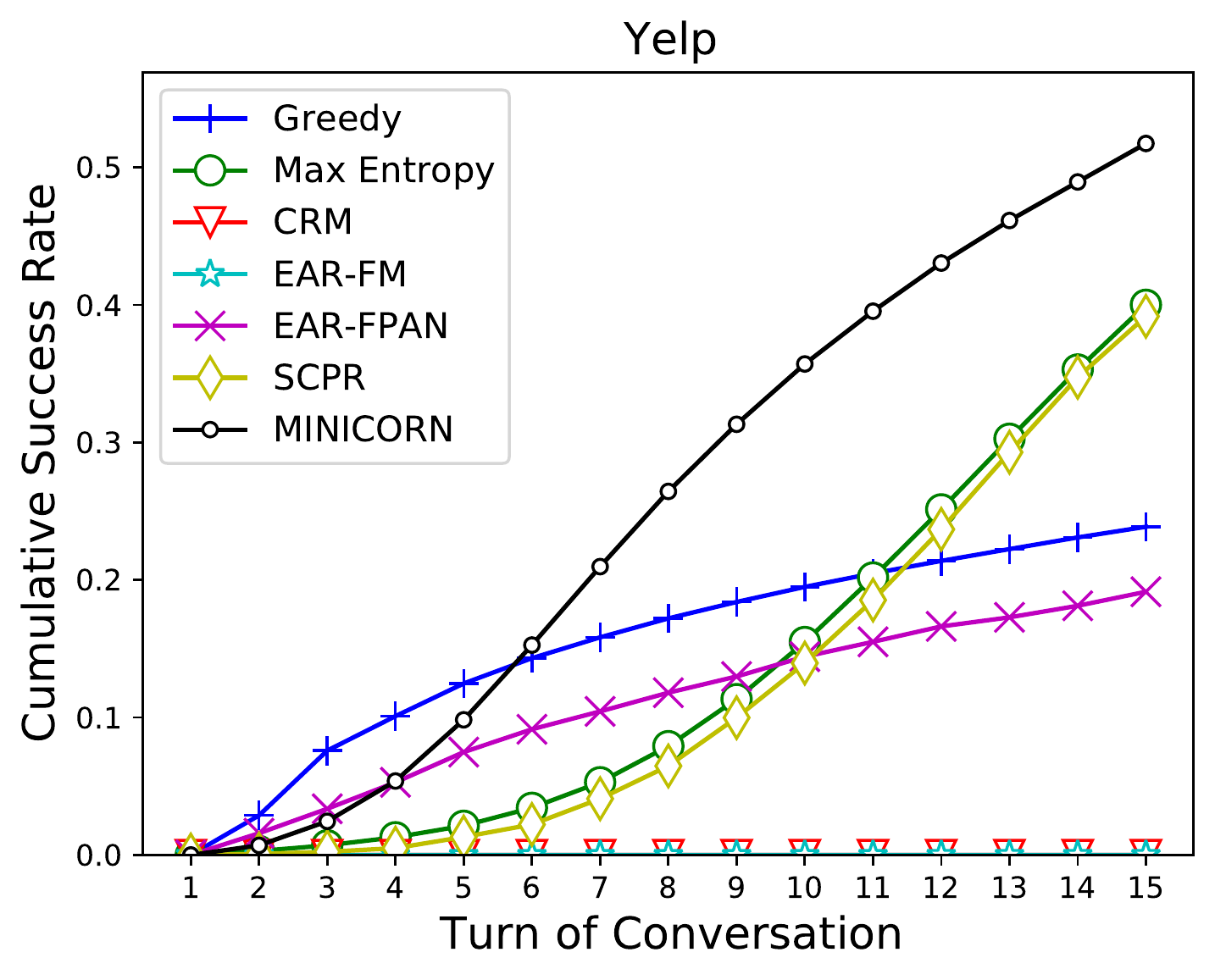}\Description[A line chart]{A line chart comparing different CRS, \modname{} achieves the best cumulative success rate from the 6$th$ turn. Greedy performs the best from the 2$^{nd}$ turn to the 5$^{th}$ turn. SCPR and Max Entropy perform comparably and both outperform Greedy from the $12^{th}$ turn.} }}%
    \caption[]{SR@T of compared baselines on \textit{LastFM} (left), \textit{MovieLens} (Middle), and \textit{Yelp} (right).}
    \label{fig:baseline}
\end{figure*}

\begin{table}[t]
\centering
\Description[A table]{A table comparing the SR@15 and AT for different CRS. On \textit{LastFM}, \modname{} outperforms the best baseline (SCPR) by 67.40\% and 13.39\% in SR@15 and AT. The relative improvements in SR@15 and AT are 7.93\% and 1.01\% on \textit{MovieLens} and 5.81\% and 3.00\% on \textit{Yelp}. }
\caption{Performance of different CRS, as measured by SR@15 and AT. The best results are in bold and the second best are underlined. ``Relative Imp.'' refers to the relative improvement of \modname{} over the best baseline.}
\label{exp:baseline}
\resizebox{\columnwidth}{!}{
\begin{tabular}{@{}lcccccc@{}}
\\ \toprule
\multirow{2}{*}{} & \multicolumn{2}{c}{\textit{LastFM}}       & \multicolumn{2}{c}{\textit{MovieLens}}   & \multicolumn{2}{c}{\textit{Yelp}}         \\\cmidrule{2-7}
                  & SR@15 ($\uparrow$)          & AT ($\downarrow$)              & SR@15 ($\uparrow$)         & AT ($\downarrow$)              & SR@15 ($\uparrow$)          & AT ($\downarrow$)             \\\midrule
                  
Greedy         & 0.218                          & 13.78                          & 0.328                          & 12.66                          & 0.239                          & {\ul 12.11}                    \\
Max Entropy          & 0.289                          & 13.88                          & 0.401                          & 12.22                          & 0.400                          & 13.41                          \\
CRM \cite{SCPR}                  & 0.325                          & 13.75                          &   --                     &         --                & 0.177                          & 13.69                          \\
CRM (reproduced)     & 0.311                          & 13.64                          & 0.019                          & 14.95                          & 0.001                          & 15.00                          \\
EAR \cite{SCPR}                  & 0.429                          & 12.88                          &   --                           &          --                      & 0.182                          & 13.63                          \\
EAR (reproduced)    & 0.435                          & 12.86                          & 0.355                          & 13.33                          & 0.001                          & 15.00                          \\
EAR-FPAN (reproduced)             & 0.431                          & 12.85                          & 0.654                          & 9.68                           & 0.191                          & 13.56                          \\
SCPR \cite{SCPR}                 & {\ul 0.465}                    & 12.86                          & --           & --           & {\ul 0.489}                    & 12.62                          \\
SCPR  (reproduced)   & 0.444                          & {\ul 12.61}                    & {\ul 0.744}                    & {\ul 8.39}                     & 0.392                          & 13.16                          \\
\modname{}              & \textbf{0.778}                 & \textbf{10.92}                 & \textbf{0.803}                 & \textbf{8.30}                  & \textbf{0.517}                 & \textbf{11.74}
 \\ \midrule
 \multicolumn{1}{c}{Ralative Imp.} & 67.40\% & 13.39\% & 7.93\% & 1.01\% & 5.81\% & 3.00\% \\
 \bottomrule
\end{tabular}
}
\end{table}
Table \ref{exp:baseline} summarizes the SR@15 and AT of different dynamic multi-shot CRS on the three datasets. 
Figure \ref{fig:baseline} shows the cumulative success rates at different conversation turns.

We make the following observations. 
First, on all three datasets, \modname{} outperforms all other baselines. One highlight is on \textit{LastFM}, \modname{} surpasses the best baseline, SCPR, by 67.40\% and 13.39\% on SR@15 and AT respectively. The small amount of training data of \emph{LastFM} may have made training difficult for RL-based approaches. The smallest margin between \modname{} and SCPR happens on \emph{MovieLens}, which has the least number of attributes and the highest data density (\# interactions $\div$ (\# users $\times$ \# items) = 3.87\%) among the three datasets. 

Second, SCPR is usually the second best in terms of SR@15 and AT. On \textit{MovieLens}, SCPR surpasses other baselines by 9.03\% in SR@15 and 13.34\% in AT. The cumulative success rates of  all CRS systems improve over time, indicating that additional rounds of preference acquisition leads to better recommendation performance.
Our reproductions of CRM, EAR, and SCPR on \textit{LastFM} perform on par or better than the reported results in \cite{EAR,SCPR}. However, on \textit{Yelp}, we were not able to exactly reproduce some reported performance, which may be due to our settings. We note that \cite{SCPR} mentions that EAR (which has the same first author with \cite{SCPR}) does not perform well on \textit{Yelp}.

The relatively high success rates of Greedy at early turns is due to the fact that Greedy makes recommendations at every turn while other methods focus on preference acquisition at early turns. 
Its lack of preference acquisition results in poor performance overall, underscoring the importance of active learning in the CRS task. 

\subsubsection{The Attribute Relation Matrix}

\begin{figure*}
    \centering
    \subfloat[\centering Learned attribute relation matrix $A$ on \textit{LastFM}. \label{fig:symA_lastfm}]{\includegraphics[width=2\columnwidth/4]{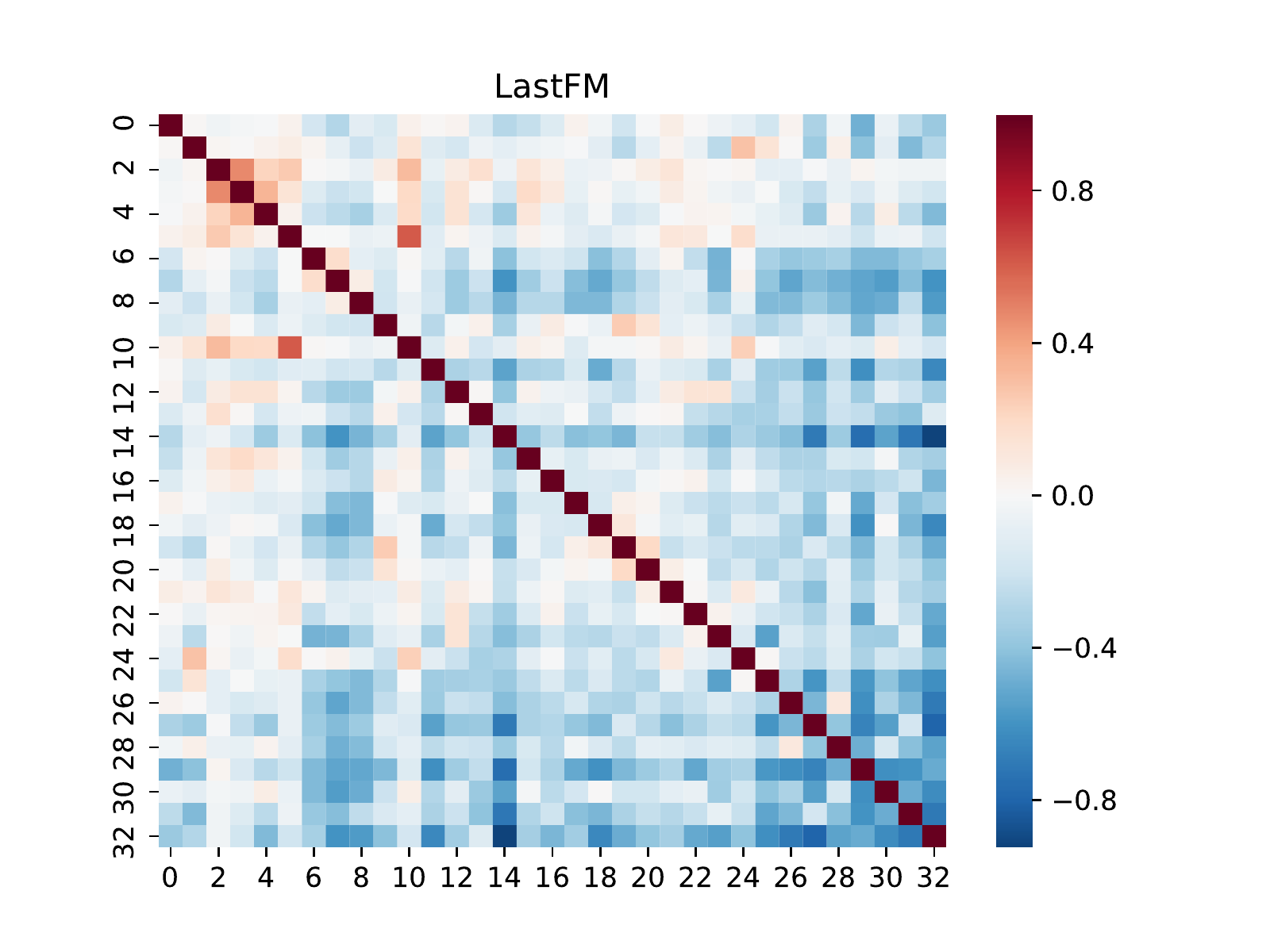} }
    \subfloat[\centering Ground-truth correlation between attributes on \textit{LastFM}. \label{fig:groundtruth_lastfm}]{\includegraphics[width=2\columnwidth/4]{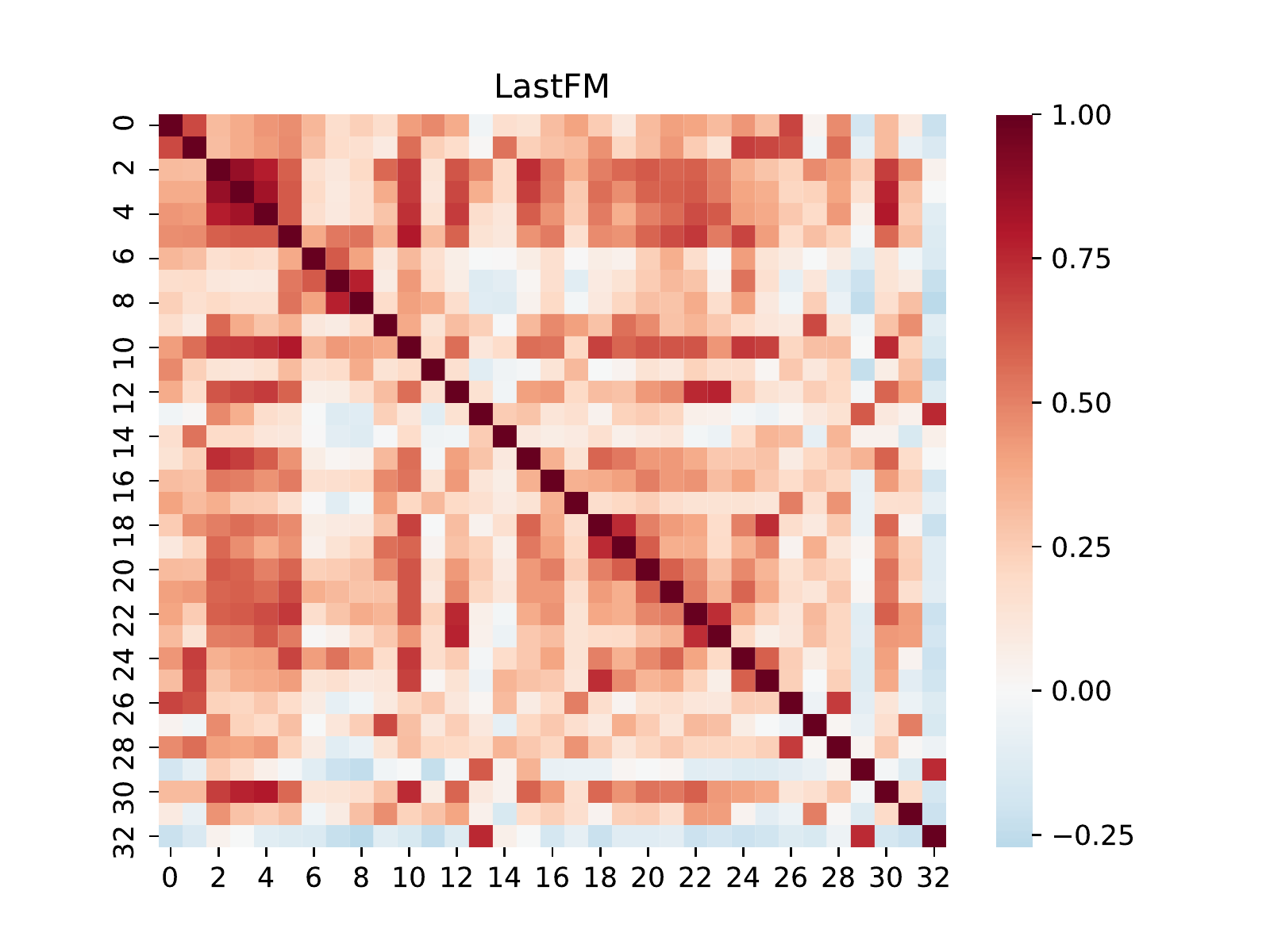} }
    \subfloat[\centering Learned attribute relation matrix $A$ on \textit{MovieLens}. \label{fig:symA_movielens}]{\includegraphics[width=2\columnwidth/4]{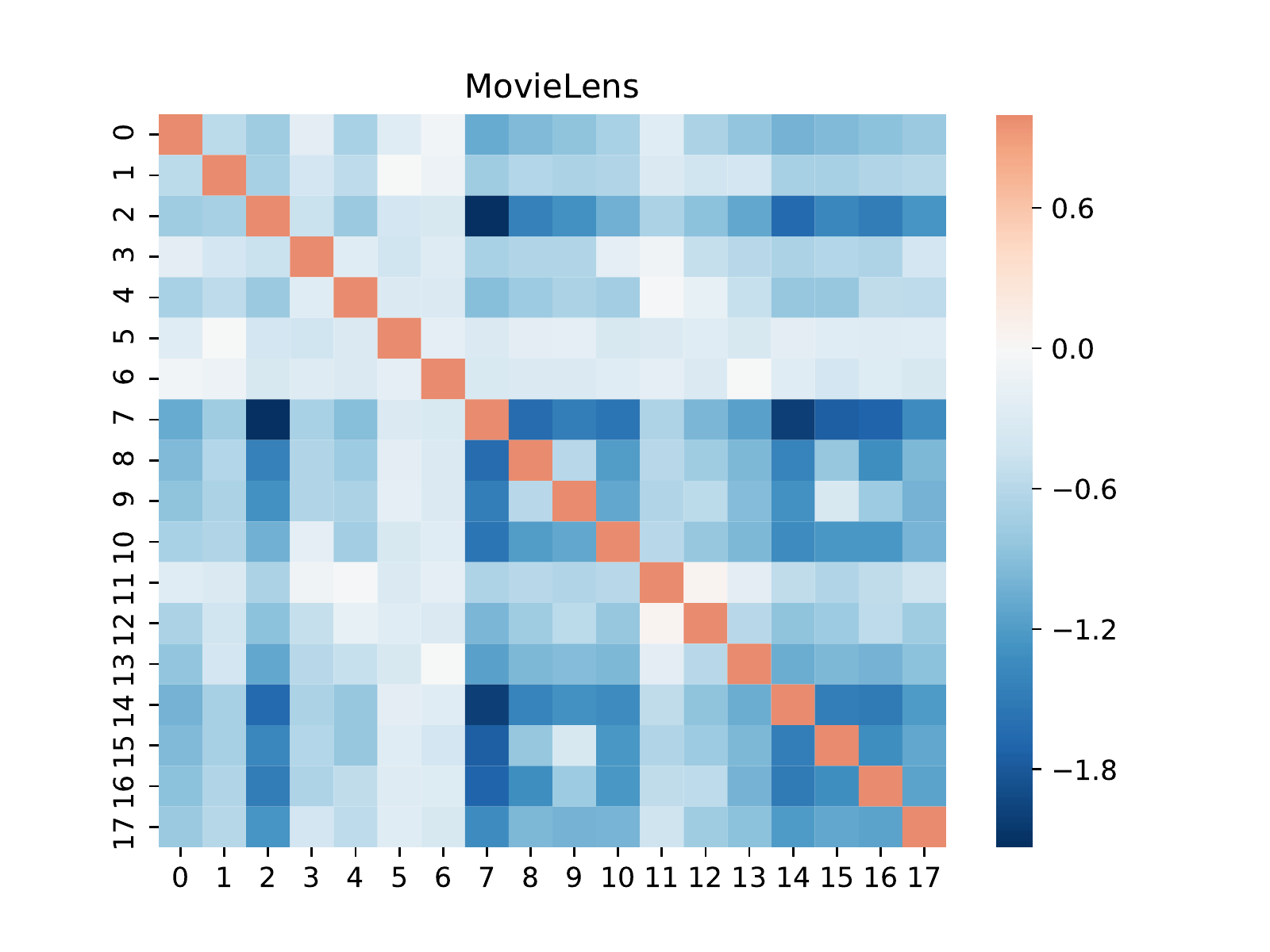} }
    \subfloat[\centering Ground-truth correlation between attributes on \textit{MovieLens}. \label{fig:groundtruth_movielens}]{\includegraphics[width=2\columnwidth/4]{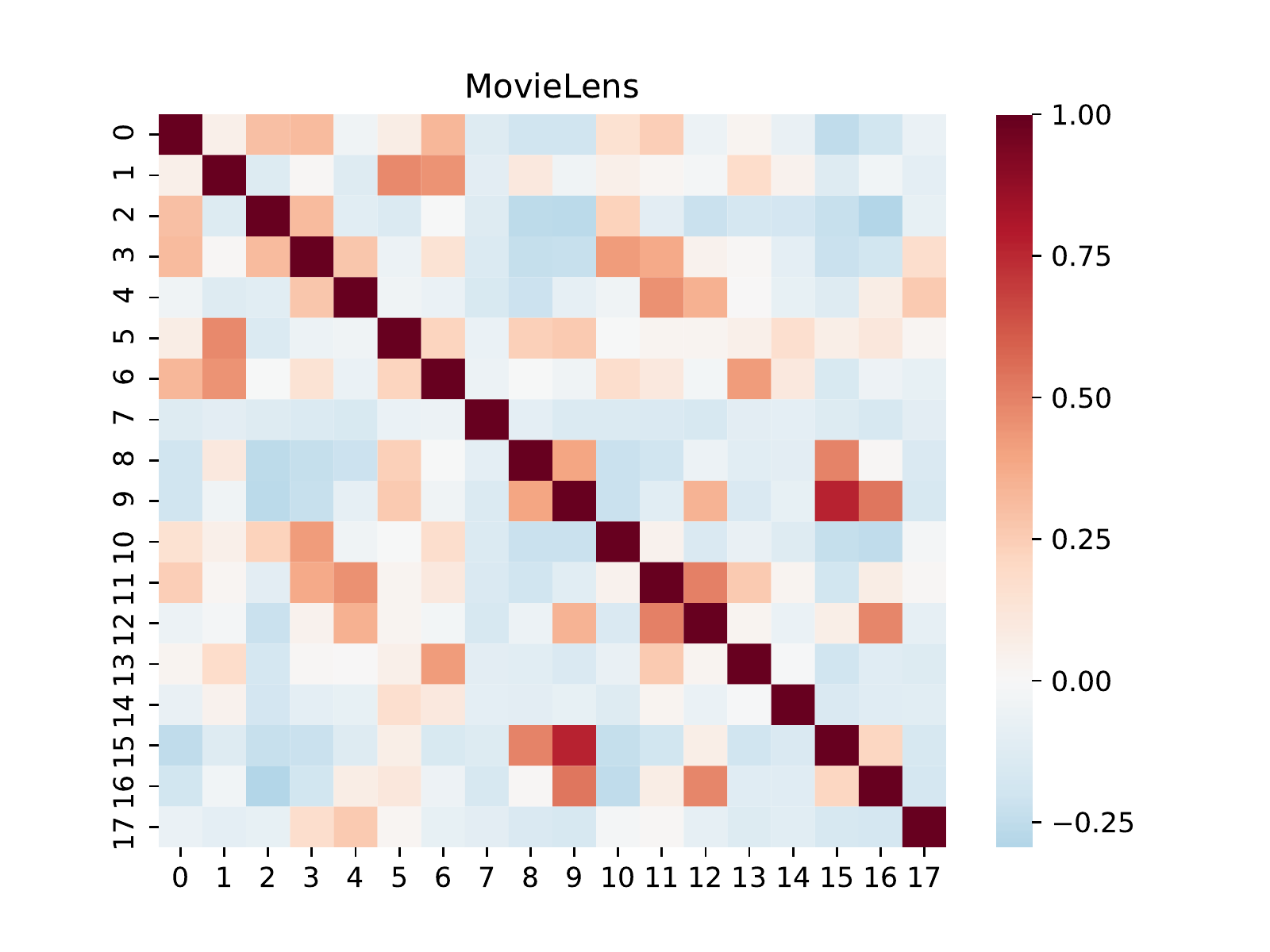} }
    \Description[Four heatmaps.]{Four heatmaps with values from -2 to 1. The diagonal values of the heatmap are all 1.}
    \caption{The Attribute Relation Matrix.}
\end{figure*}

We visualize the learned attribute relation matrix $A$ in Figure \ref{fig:symA_lastfm} and Figure \ref{fig:symA_movielens} from \textit{LastFM} and \textit{MovieLens}, respectively. Figure \ref{fig:groundtruth_lastfm} and Figure \ref{fig:groundtruth_movielens} show the ground-truth correlation matrix between attributes.

We observe close resemblance between the two matrices on both datasets, which appear to be mean-shifted versions of each other. 
First, the two matrices share similar cluster structures, such as the cluster of attributes 0-6 on \textit{LastFM}.
Second, we find more popular attributes are usually more positively correlated to the other attributes in both matrices. For example, on \textit{MovieLens}, the 6$^\text{th}$ (\emph{Drama}) attribute and the 5$^\text{th}$ (\emph{Comedy}) attribute in \textit{MovieLens} are positively correlated to most attributes and they are the most popular ones associated with 1413 (39.8\%) and 1126 (31.7\%) items, respectively.
Third, on \textit{MovieLens}, we find the 11$^\text{th}$ attribute (\emph{Action}) and 12$^\text{th}$ attribute (\emph{Adventure}) have the highest correlation score of 0.06, which is in line with the original dataset, where the correlation between the two attributes is 0.50.
Among the 276 \emph{Adventure} movies, 126 (45\%) movies have the attribute \emph{Action}.

Treating one matrix as a vector, we compute the correlation between the two matrices, which is 0.82 on \textit{LastFM} and 0.79 on \textit{MovieLens}. This demonstrates that the training procedure allows \modname{} to recover correlations among attributes.

\subsection{Computational Efficiency}
To evaluate data efficiency and wall-clock running time, we compare \modname{} with SCPR, the state-of-the-art, best performing RL baseline on the LastFM dataset with two Intel Xeon 4114 CPUs.

\begin{figure}
    \centering
    {\includegraphics[width=\columnwidth]{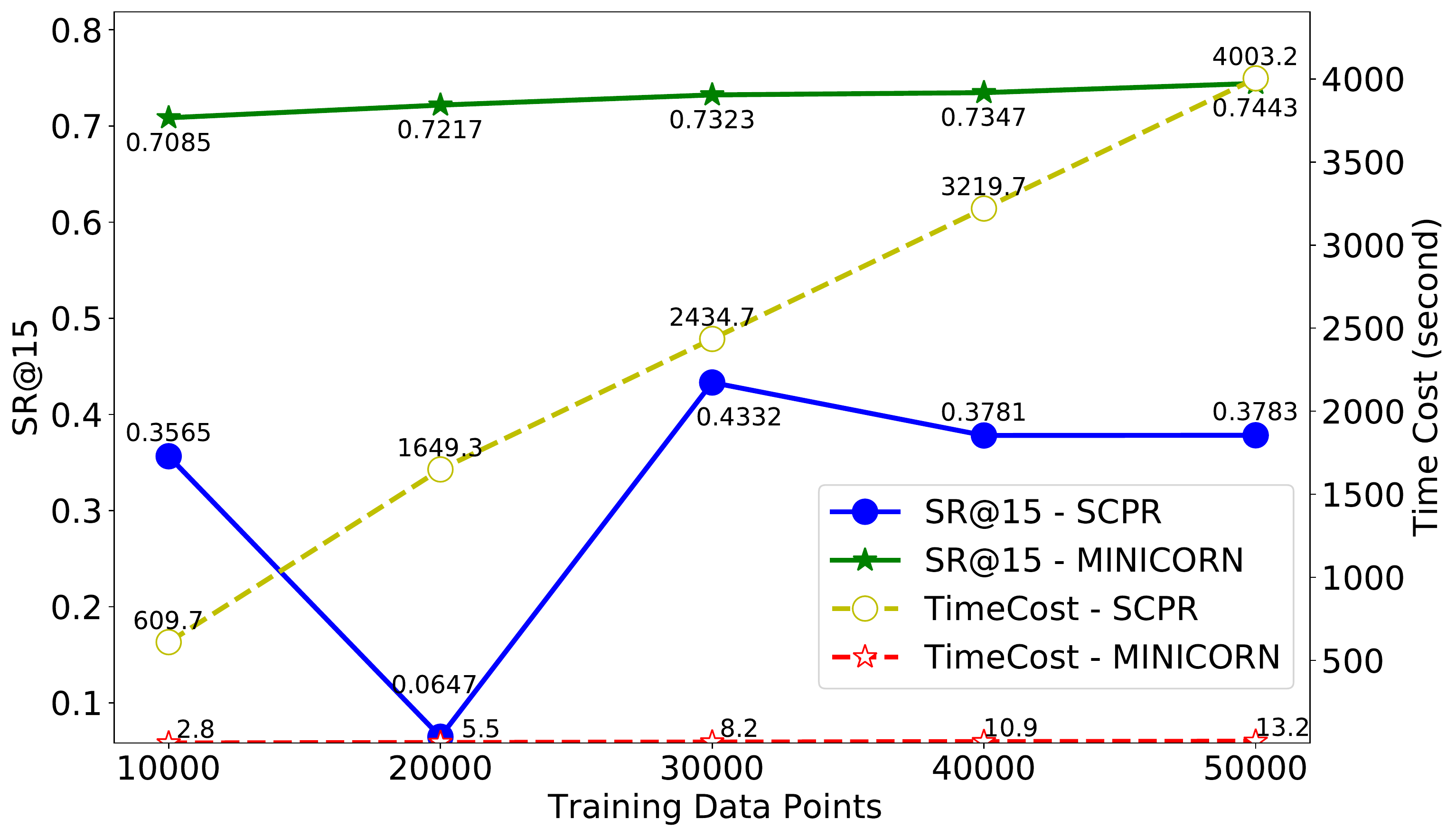}}
    \Description[A line chart]{A line chart. The SR@15 of \modname{} varies from 0.7085 to 0.7443, and the SR@15 of SCPR varies from 0.0647 to 0.4332. The time costs are 2.8, 5.5, 8.2, 10.9, 13.2 seconds for \modname{} to train from 10000 to 50000 data points. For SCPR, it costs 609.7, 1649.3, 2434.7, 3219.7, and 4003.2 seconds.}
    \caption{Computational efficiency, measured by SR@15 and wall-clock time with equal number of training data points on \textit{LastFM}.}
    \label{exp:computation}
\end{figure}

For data efficiency, we compare model performance after equal number of data points have been used in stochastic gradient descent. The same data point used in two epochs is counted as two data points. 
Figure \ref{exp:computation} indicates \modname{} achieves substantially better performance under equal number of data points. Also, low performance of SCPR at 20,000 data points suggests issues with the stability of training.

Second, Figure \ref{exp:computation} shows the wall-clock time of training both SCPR and \modname{} with different training data points. For SCPR, it takes more than 1 hour for training 50,000 data points. For \modname{}, it takes 13.2 seconds. 
Together, the results demonstrate that \modname{} provides significant computational cost reduction over Reinforcement Learning.
\subsection{Ablation Studies}
\begin{figure*}
    \centering
    \subfloat[\centering SR@T of different attribute selection strategies. \label{fig:MovieLens_Selection}]{\includegraphics[width=2\columnwidth/3]
    {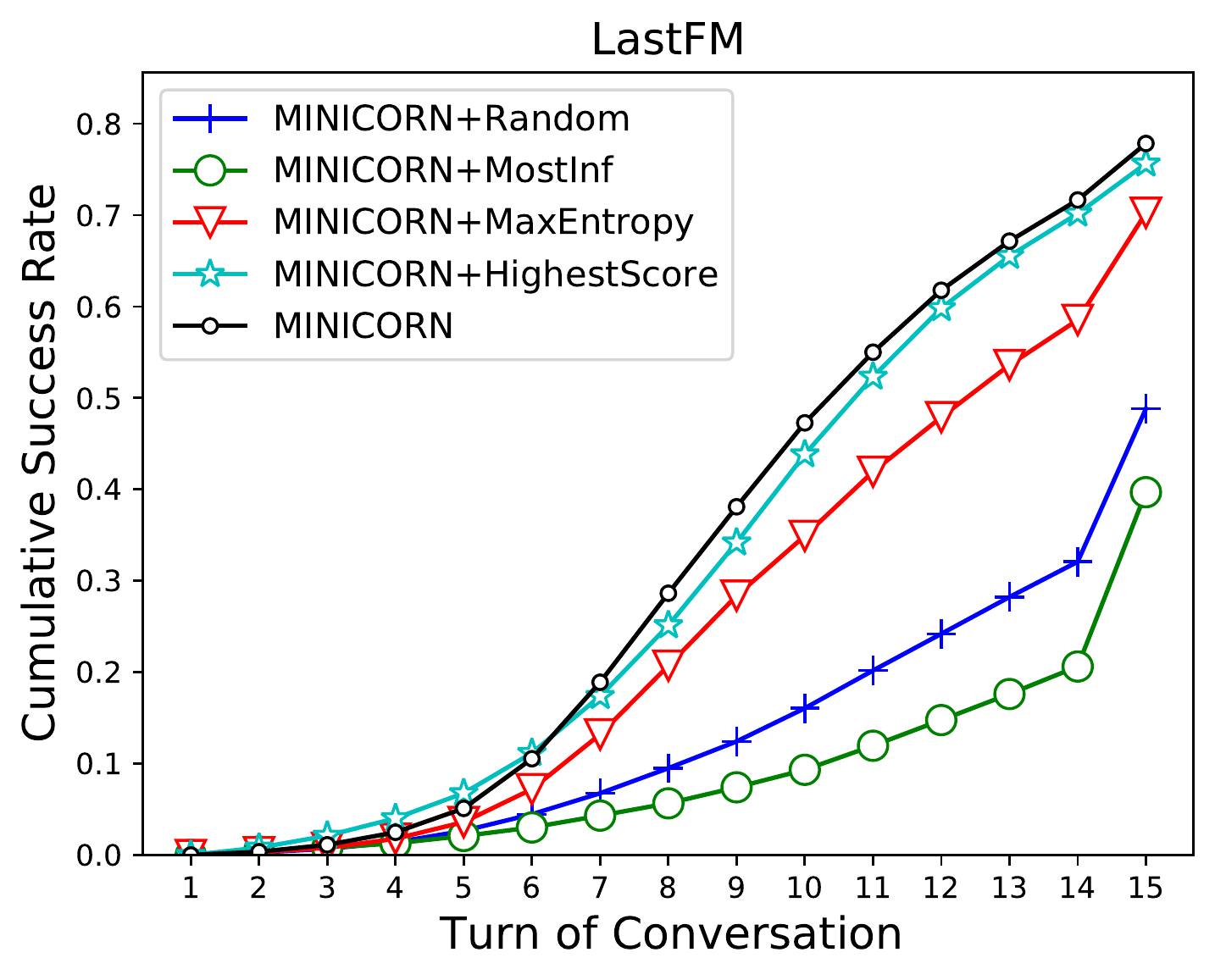}\Description[A line chart.]{A line chart. The original \modname{} outperforms the best baseline (\modname{}+Highest Score) by about 0.02 in SR after the 8$^{th}$ turn. \modname{}+MostInf performs the worst.}}
    \subfloat[\centering $|\mathcal{V}_t|$ of different attribute selection strategies. \label{fig:MovieLens_CandidateItemSize}]{\includegraphics[width=2\columnwidth/3]
    {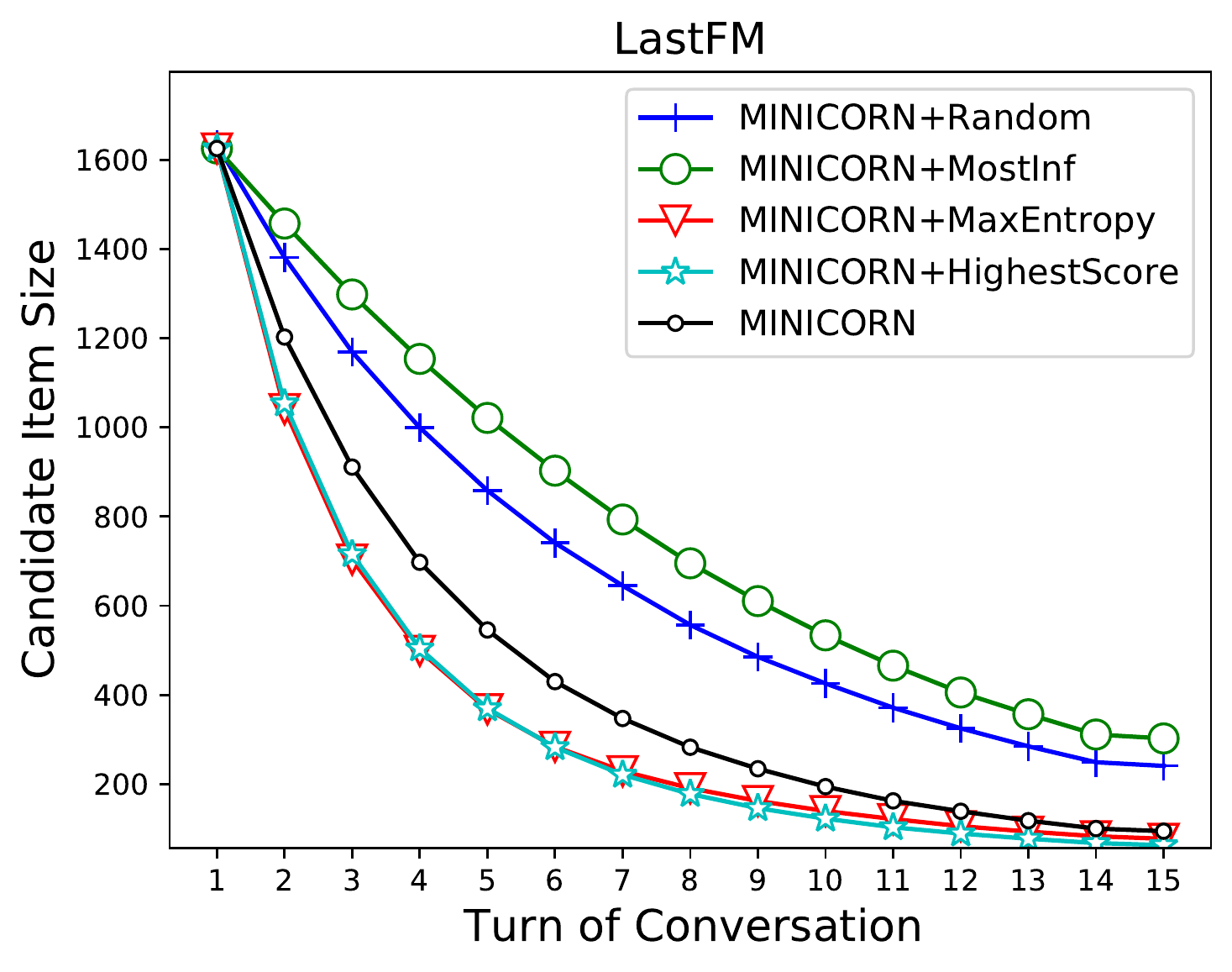}\Description[A line chart.]{A line chart. \modname{}+MaxEntropy and \modname{}+HighestScore reduce the candidate item set the most, followed by the original \modname{}.}} 
    \subfloat[\centering SR@T of different attribute preference vectors $\bm{q}$. \label{fig:attribute_inf}]{\includegraphics[width=2\columnwidth/3]{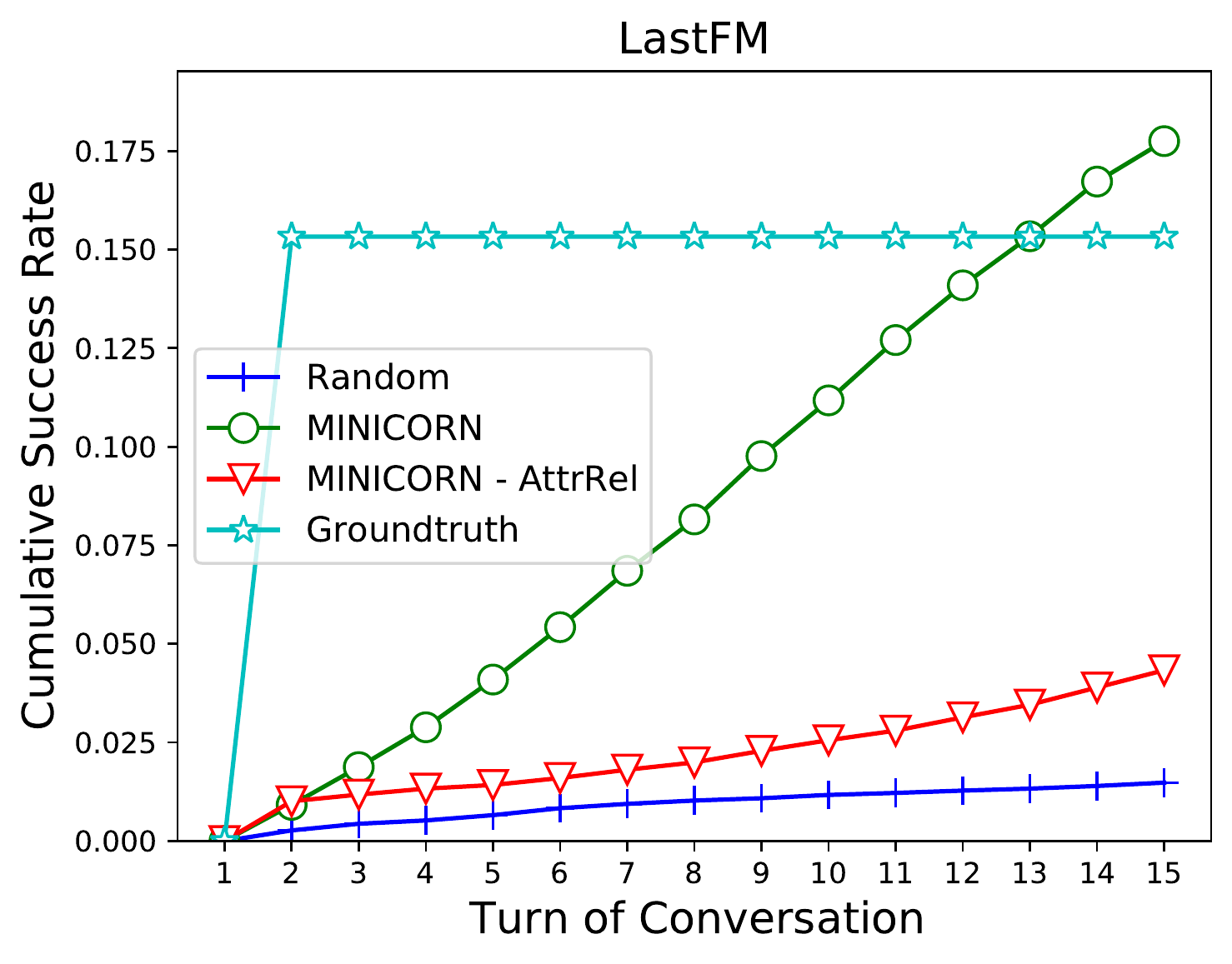}\Description[A line chart.]{A line chart. \emph{Groundtruth} achieves the SR of 0.15 from the $2^{nd}$ turn. The original \modname{} outperforms the \emph{Groundtruth} variation from the 13$^{th}$ turn. \emph{Random} performs the worst. }}\\
    \subfloat[\centering SR@T of different recommendation methods.\label{fig:LastFM_Recommendation}]{\includegraphics[width=2\columnwidth/3]{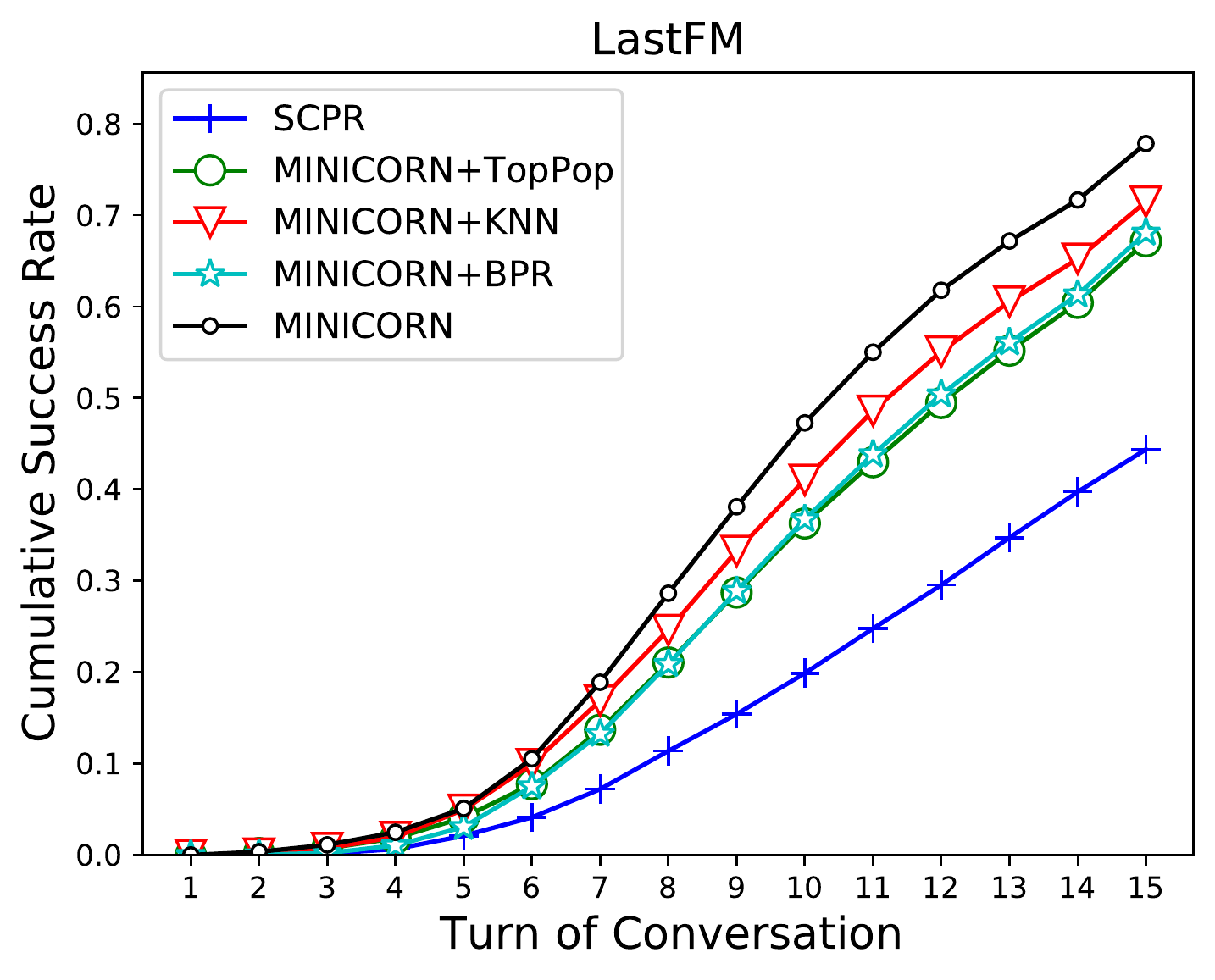}\Description[A line chart.]{A line chart. \modname{} performs the best, followed by the variation of \modname{}+KNN. \modname{}+TopPop and \modname{}+BPR receive comparable results. SCPR performs the worst.}}
    \subfloat[\centering SR@T of different confidence threshold $\alpha$. \label{fig:hyperparameters}]{\includegraphics[width=2\columnwidth/3]{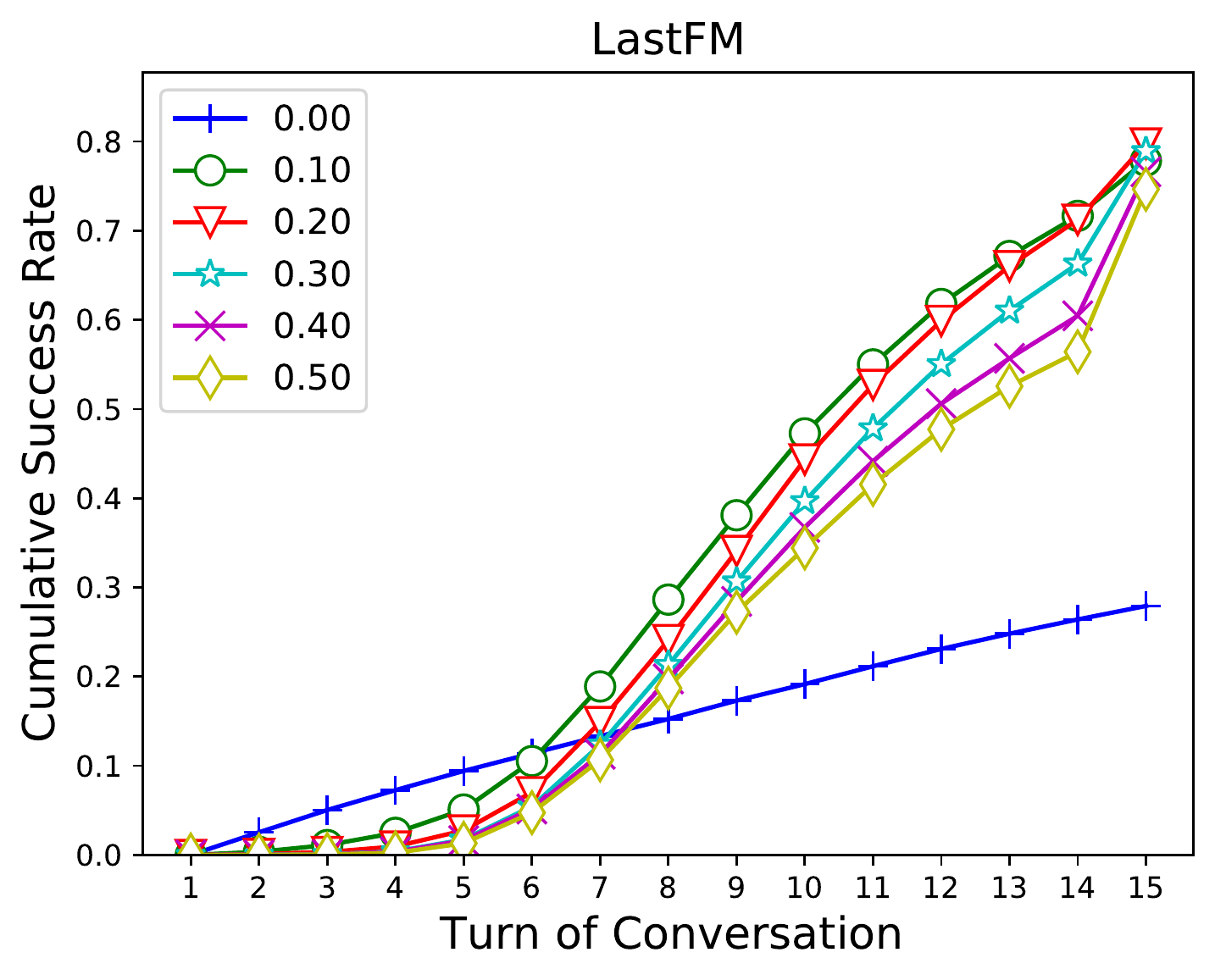}\Description[A line chart.]{A line chart. When $\alpha=0.0$, it performs the best before the 6$^{th}$ turn and the worst after the 8$^{th}$ turn. The performances are comparable when setting $\alpha$ from 0.1 to 0.5. }}
    \subfloat[\centering Case studies comparing \modname{} and SCPR with the sizes of the candidate sets. 
    \label{fig:casestudy}]{\includegraphics[width=2\columnwidth/3]{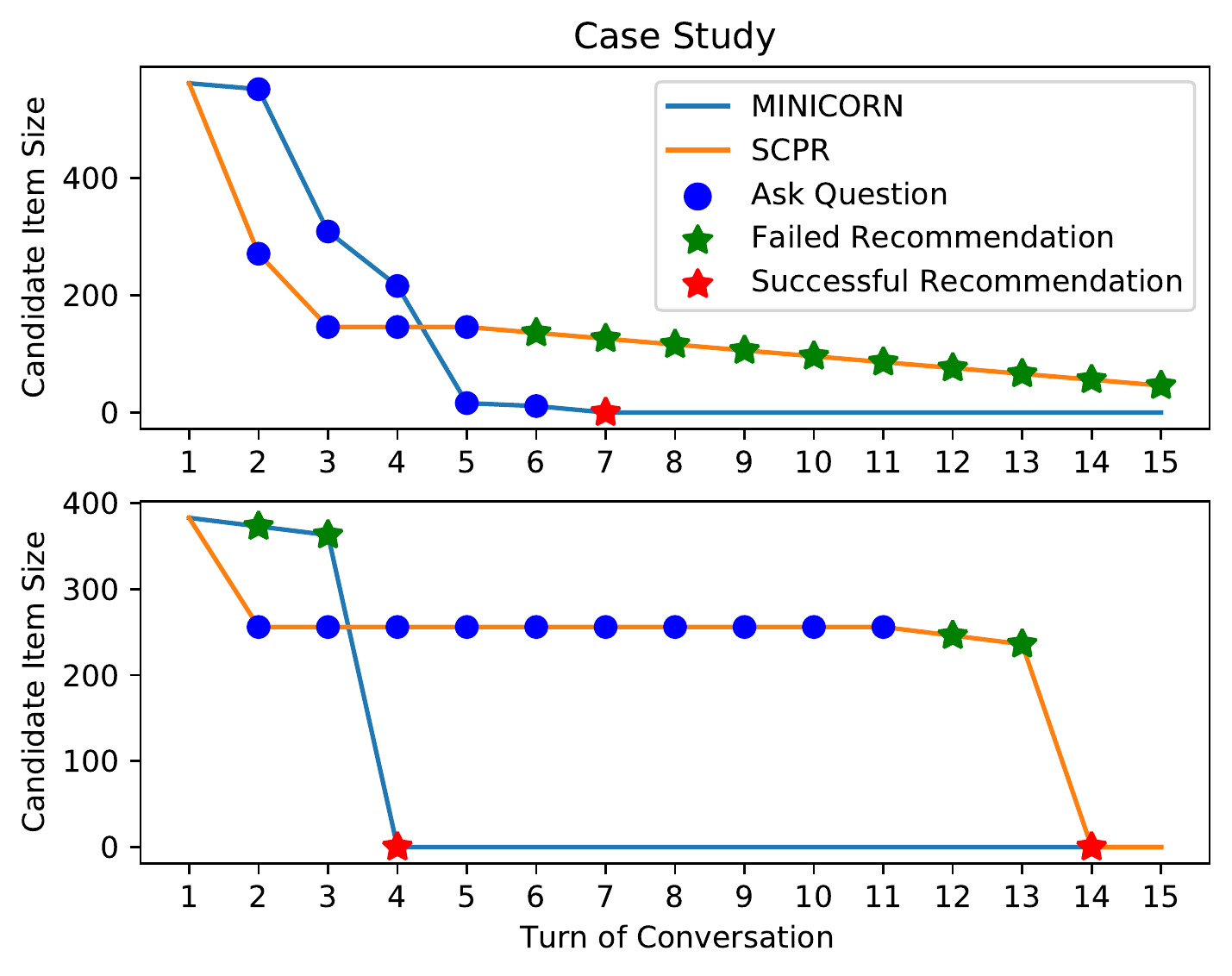}\Description[Two line charts.]{Two line charts. In the first case, both methods ask questions in the beginning. Then \modname{} recommend successfully in the 7$^{th}$, where there are only 11 candidate items. SCPR chooses to recommend from the 6$^{th}$ turn and does not succeed until the end. In the second case, \modname{} recommend from the 2$^{nd}$ turn (383 candidate items) and succeed in the 4$^{th}$ turn. SCPR starts recommendation in the 12$^{th}$ turn and succeeds in the 14$^{th}$ turn.  } }
    \caption{Ablation Study on \textit{LastFM}.}
    \label{fig:ablation}
\end{figure*}

To evaluate the effectiveness of each component in \modname{}, we perform a series of ablation studies on \textit{LastFM}.
We received similar results on \textit{MovieLens}. Due to the page limitation, we report the results of \textit{MovieLens} in the supplemental material.
\subsubsection{Comparison of Attribute Selection Strategies}
\label{sec:attribute_select}
In this experiment, we evaluate the efficiency of attribute selection strategies, which determine which attribute that the system should query the user about.
We replace the original uncertainty-based selection strategy in \modname{} with the following:
\begin{itemize}
    \item \emph{Random}, which randomly selects an attribute that has not been queried.
    \item \emph{MostInf}, which selects the attribute with the highest potential to alter the recommended item list. When the CRS queries the user about an attribute, the answer will be either positive or negative. Feeding the two possible user feedback vectors (i.e., $a_{t,p}$ set to 0 or 1) to \modname{} yields two lists of recommended items. We select the attribute leading to the least intersection between the two lists, as this attribute exerts the most influence on recommendation.
    \item \emph{MaxEntropy}, which is similar to the MaxEntropy strategy in Section \ref{baseline}. It selects the attribute that, in expectation, reduces the candidate set the most.
    \item \emph{HighestScore}, which selects the attribute that the user is most likely to have positive preference. It is adopted by several existing CRS methods \cite{QandA,yongeng2018,EAR,KBQG}.
\end{itemize}

Figure \ref{fig:MovieLens_Selection} shows the cumulative success rate and Figure \ref{fig:MovieLens_CandidateItemSize} shows the number of remaining candidate items, $|\mathcal{V}_t|$, averaged over all users, at every conversational turn. 

The proposed uncertainty-based strategy outperforms all baseline methods, including the widely used \emph{HighestScore} strategy.
As we expect, the \emph{MaxEntropy} strategy leads to maximum reduction of the candidate set. \emph{HighestScore} snatches the second spot in success rate with the second fastest reduction. Interestingly, \modname{} achieves the best success rate but is only the third fastest in reduction of the candidate set. This indicates that the strength of \modname{} extends beyond merely candidate elimination. We investigate the cause of this phenomenon in the next ablation study.

\subsubsection{Comparison of Preference Estimation Strategies}
\label{sec:attribute_embedding}
In this ablation study, we empirically verify the preference estimation strategy used in \modname{} and investigate its effect on the Recommendation Network (RN). In \modname{}, we estimate user preference $\bm{q}_t$ using the attribute correlation matrix $A$ as $\bm{q}_t = A \bm{a}_t$, where $\bm{a}_t$ is the aggregated user feedback. 

We create three ablation baselines. The first one, \emph{Random}, sets $\bm{q}_t$ to a random vector. The second one, \emph{\modname{} - AttrRel}, removes $A$ and sets $\bm{q}_t$ directly to the user feedback vector $\bm{a}_t$. The final one, \emph{GroundTruth}, sets $\bm{q}_t$ to the ground-truth attribute vector of the target item $\bm{b}(v)$. 

To isolate the effects $\bm{q}_t$ on RN from the overall conversation strategy, we make two adjustments to \modname{}. First, we do not update the candidate item set after receiving user feedback. That is, RN chooses from all possible items throughout the conversation turns. Second, at every turn of the conversation, we let \modname{} ask one question and make recommendations. 

Figure \ref{fig:attribute_inf} shows the results. \modname{} outperforms the \emph{Random} and the \emph{\modname{} - AttrRel} ablations; \emph{\modname{} - AttrRel} outperforms \emph{Random}.
From these results, we make two conclusions. First, the attribute relation matrix $A$ contributes significantly to the performance of RN. Second, the benefits of the acquired attribute preferences go beyond simple elimination of candidate items.

Surprisingly, \modname{} performs better than the \emph{GroundTruth} baseline at conversational turns 14 and 15. We hypothesize that using the ground-truth attributes for training leads to slight overfitting and degraded test performance.

\subsubsection{Comparison of Recommendation Methods}
\label{sec:recexp}
To evaluate the effectiveness of the Recommendation Network (RN) in \modname{}, we compare it to a set of strong traditional recommendation methods:
including the non-personalized method, TopPop, which always recommends the most popular item. We also include the strong K-nearest-neighbor (KNN) method \cite{itemknn,recsys19} and matrix factorization method BPR \cite{bpr}.
To perform these traditional methods under the multi-shot conversational recommendation setting, we replace the original RN in \modname{}. That is, the user feedback on attribute preferences only updates the candidate items while not influencing the recommendation policy. 

Figure \ref{fig:LastFM_Recommendation} show the SR@T on \textit{LastFM}.
We include SCPR method in this comparison since it is the best baseline method. We have the following observations. First, the superiority of \modname{} compared to those strong traditional recommendation methods indicates the effectiveness of the proposed NN. Second, the traditional methods with the selection policy in \modname{} outperform the best multi-shot CRS baseline, which suggests the advantage of the Conversation Decision Module in \modname{}.

\subsubsection{Sensitivity of Threshold $\alpha$}
We analyze the sensitivity of the confidence threshold $\alpha$. As shown in Figure 
\ref{fig:hyperparameters}, we vary the threshold from $0.0$ to $0.5$. The threshold $\alpha = 0.0$ refers to regarding all the attributes as certain ones and making recommendations throughout the conversation turns. On the other hand, $\alpha = 0.5$ represents that only those attributes that are assigned with 1 or 0 in the predicted belief embedding $\bm{q}_t$ are certain ones.

Similar to Greedy in Figure \ref{fig:baseline}, the performances of $\alpha = 0.0$ achieves a relatively high success rate at early conversation turns ($t \leq 6$) but fails later due to the limited information gain. 
On \textit{LastFM}, $\alpha = 0.1$ receive the best performances in general.

\subsubsection{Case Study comparing SCPR and \modname{}}
We investigate the difference between \modname{} and SCPR through two case studies from the \emph{LastFM} test set.
In Figure \ref{fig:casestudy}, we visualize the candidate itemset size and the actions (making query or recommendation) according to the conversational turns.

We have the following observations. Compared to \modname{}, recommendations in SCPR usually have worse timing. In the first case (Figure \ref{fig:casestudy}, top), SCPR starts making recommendation early, which leads to a series of failed recommendations. In the second case (Figure \ref{fig:casestudy}, bottom), SCPR keeps asking questions until the 12$^{th}$ turn, but the questions do not reduce the candidate set significantly. Upon close inspection, we realize that SCPR does not use negative feedbacks to attribute queries to update the candidate set or the recommendation module. This results in general inefficiency.

\section{Conclusion}

In this paper, we propose a novel non-RL method, \modname{} for the multi-shot conversational recommendation. 
\modname{} utilizes a simple rule to decide whether to ask questions or to recommend and query the users for the attribute with the highest uncertainty.
Extensive experimental results on three real-world datasets demonstrate the effectiveness of the proposed \modname{}, where \modname{} achieves a higher success rate within fewer conversational turns.


\bibliographystyle{ACM-Reference-Format}
\bibliography{sample-base}

\appendix
\appendix
\setcounter{page}{1}
\section{Notation Table}
\begin{table}[H]
    \renewcommand{\arraystretch}{1.2}
\centering
\caption{Frequently used notations}
\begin{tabular}{lp{18.5em}}
\toprule
$u$, $v$, $p$ & a user, an item, and an attribute \\
$\mathbf{e}^{\text{user}}$, $\mathbf{e}^{\text{item}}$ & user embedding, item embedding \\
$\mathbf{e}^{\text{attr}}$ & attribute embedding\\
$\mathbf{b}(v) \in \{0,1\}^P$ & binary ground-truth attributes for item $v$\\
$\mathcal{H}_u$ & representative history of user $u$ \\
$\mathcal{V}[p]$ & set of items with the attribute $p$ \\
$\mathcal{P}_v$ & set of attributes associated with item $v$ \\
$\mathcal{V}_t$ & set of candidate items at the $t^{\text{th}}$ turn \\
$\mathbf{a}_t \in \{0,0.5,1\}^P$ & cumulative user feedback by the $t^{\text{th}}$ turn \\
$\mathbf{q}_t \in [0,1]^P$ & predicted attribute preferences at the $t^{\text{th}}$ turn \\
$K$ & number of items to recommend at one turn \\

\bottomrule
\end{tabular}
\end{table}

\section{Ablation Studies on \textit{MovieLens}}
\label{app:movielens}
We also did experiments to compare different selection strategies and different recommendation methods on \textit{MovieLens}.

\begin{figure*}[t!]
    \centering
    \subfloat[\centering SR@T of different attribute selection strategies. \label{fig:ml_Selection}]{\includegraphics[width=2\columnwidth/3]{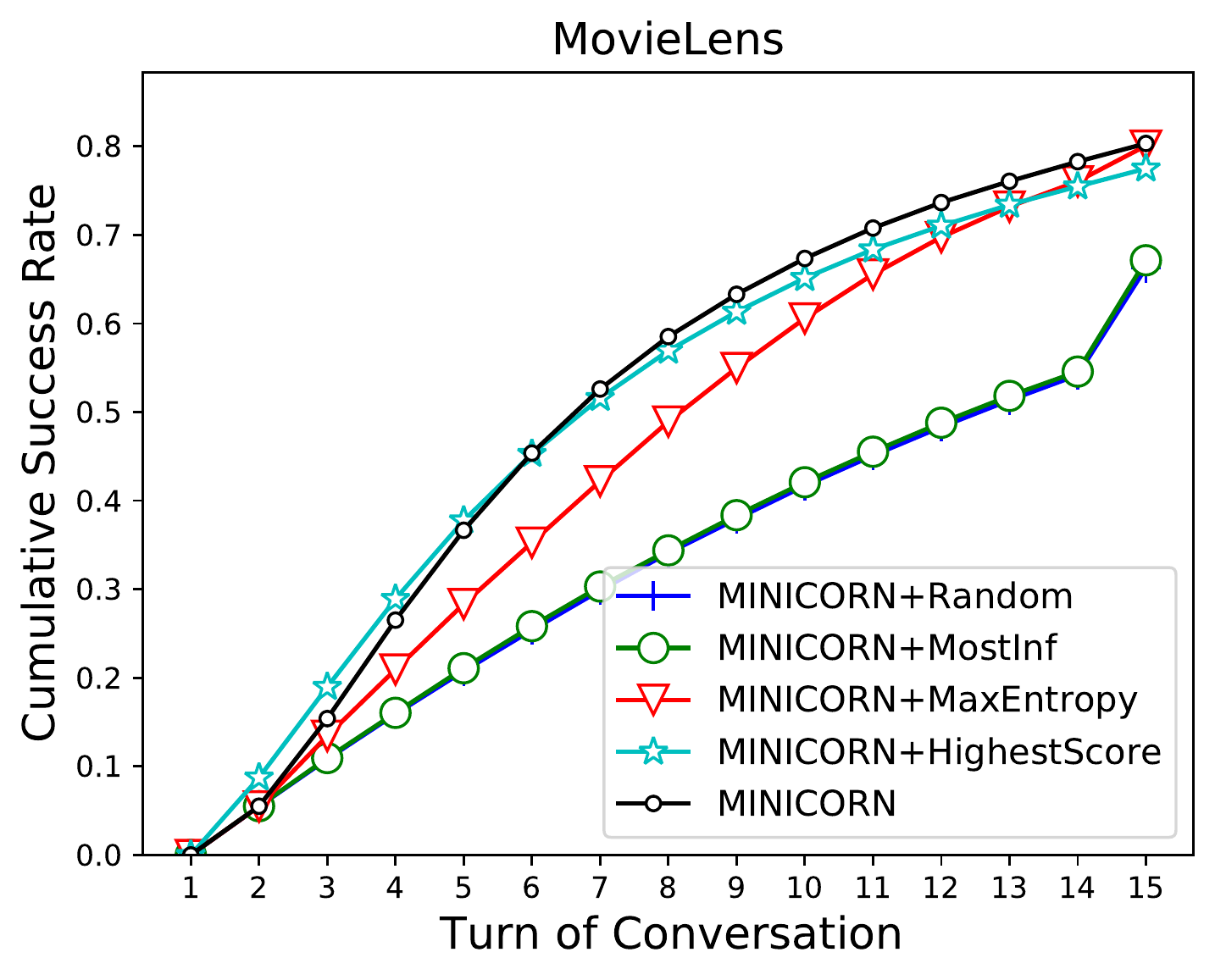} }
    \subfloat[\centering $|\mathcal{V}_t|$ of different attribute selection strategies. \label{fig:ml_CandidateItemSize}]{\includegraphics[width=2\columnwidth/3]{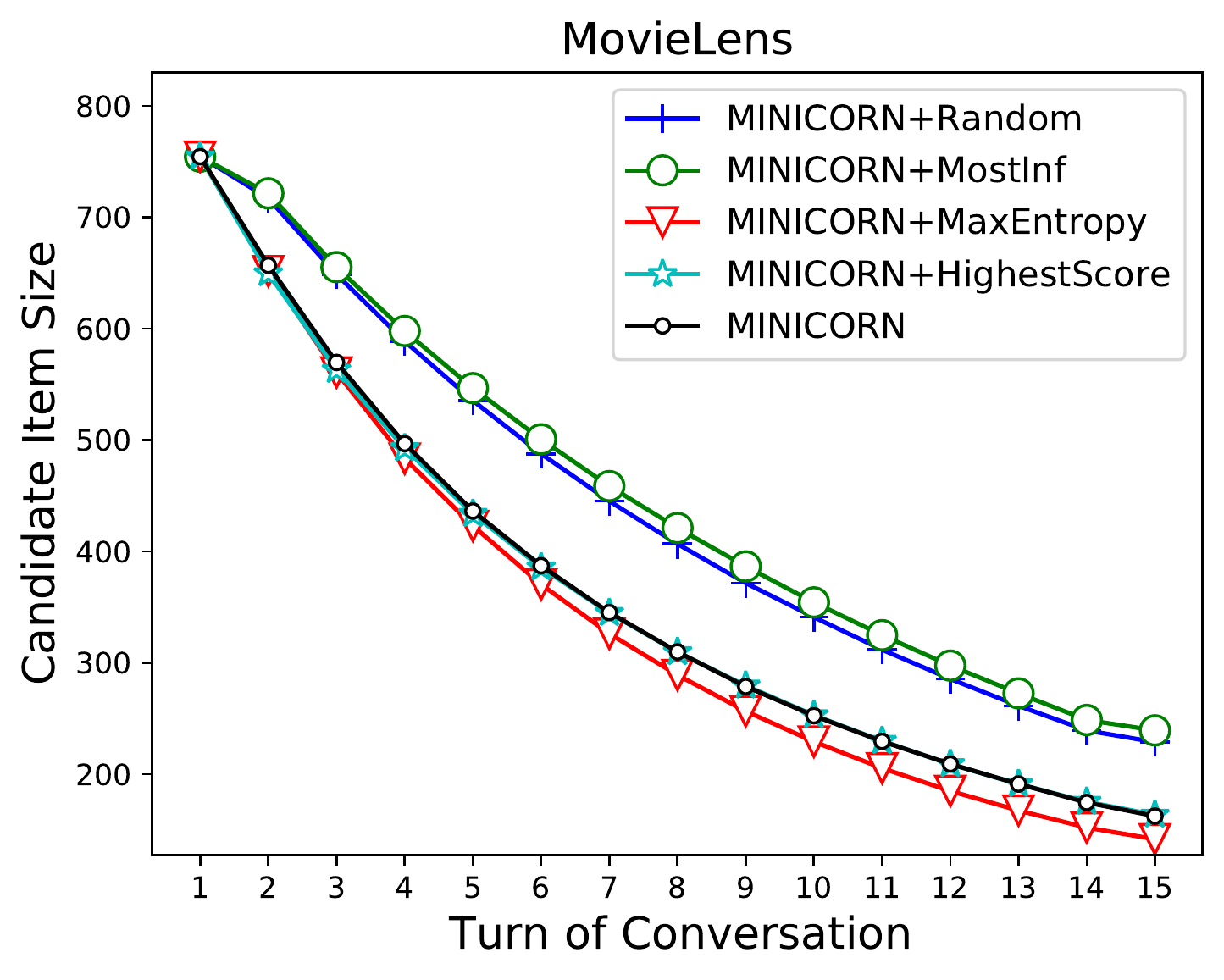} }
    \subfloat[\centering SR@T of different recommendation methods.
    \label{fig:ml_Recommendation}]{\includegraphics[width=2\columnwidth/3]{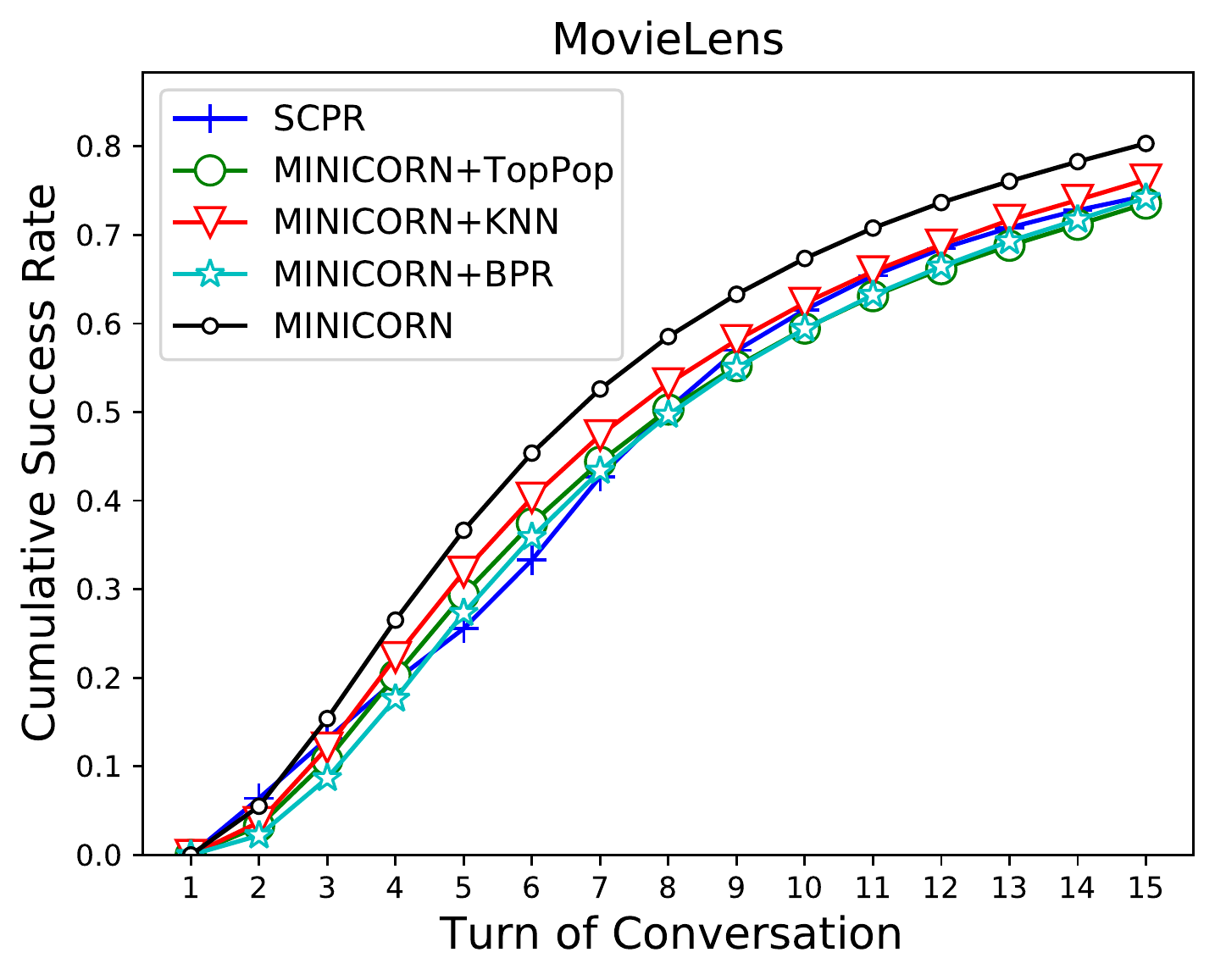}}
    
    \caption{Ablation Study on \textit{MovieLens}.}
    \label{fig:ml_ablation}
\end{figure*}

Table \ref{exp:selection} summarizes the SR@15 and AT of using different recommendation methods on \textit{LastFM} and \textit{MovieLens}. 
Figure \ref{fig:ml_Selection} shows the cumulative success rate and Figure \ref{fig:ml_CandidateItemSize} shows the number of remaining candidate items, $|\mathcal{V}_t|$, averaged over all users, at every conversational turn on \textit{MovieLens}. 
We receive the same results in Section \ref{sec:attribute_select}, where the proposed \modname{} outperforms all other attribute selection strategies. 

\begin{table}[H]
\centering
\caption{Performances of attribute selection strategies. The best results are in bold and the second best are underlined. ``Relative Imp.'' refers to the relative improvement of \modname{} over the best baseline.
}
\label{exp:selection}
\begin{tabular}{@{}ccccc@{}}
\\ \toprule

\multirow{2}{*}{}     & \multicolumn{2}{c}{\textit{LastFM}}                                              & \multicolumn{2}{c}{\textit{MovieLens}}                                          \\ \cmidrule{2-5}
                      & \multicolumn{1}{c}{SR@15($\uparrow$)} & AT($\downarrow$)              &\multicolumn{1}{c}{SR@15($\uparrow$)} & AT($\downarrow$)             \\ \midrule
\modname{}+Random       & 0.488          & 13.412          & 0.663          & 10.791          \\
\modname{}+MostInf      & 0.397          & 14.009          & 0.671          & 10.747       \\
\modname{}+MaxEntropy   & 0.702          & 11.872          & {\ul 0.801}    & 9.053          \\
\modname{}+HighestScore & {\ul 0.756}    & {\ul 11.068}    & 0.775          & {\ul 8.372}    \\
\modname{}              & \textbf{0.778} & \textbf{10.919} & \textbf{0.803} & \textbf{8.300}
 \\ \midrule
Relative Imp.        & 2.91\%         & 1.35\%          & 0.25\%         & 0.86\%         \\
\bottomrule
\end{tabular}
\end{table}

\begin{table}[H]
\centering
\caption{Performances of using different recommendation methods. The best initialization results of each model are in bold faces and the second best results are underlined. ``Relative Imp.'' refers to the relative improvement of \modname{} over the best recommendation method.
}
\label{exp:recommendation}
\begin{tabular}{@{}ccccc@{}}
\\ \toprule

\multirow{2}{*}{}     & \multicolumn{2}{c}{LastFM}                                              & \multicolumn{2}{c}{MovieLens}                                          \\ \cmidrule{2-5}
                      & \multicolumn{1}{c}{SR@15($\uparrow$)} & AT($\downarrow$)              &\multicolumn{1}{c}{SR@15($\uparrow$)} & AT($\downarrow$)             \\ \midrule
SCPR                                  & 0.444                     & 12.608          & 0.744                     & {\ul 8.385}    \\
\modname{}+TopPop                                & 0.671          & 11.771          & 0.735          & 9.205 \\
\modname{}+KNN                                   & {\ul 0.715}    & {\ul 11.370}    & {\ul 0.762}    & 8.880       \\
\modname{}+BPR                                   & 0.681          & 11.767          & 0.742          & 9.300        \\
\modname{}                               & \textbf{0.778} & \textbf{10.919} & \textbf{0.803} & \textbf{8.300}
 \\ \midrule
 Relative Imp. & 8.91\%         & 3.97\%          & 5.38\%         & 1.01\%  \\
 \bottomrule
\end{tabular}
\end{table}

Table \ref{exp:recommendation} summarizes the SR@15 and AT of using different recommendation methods on \textit{LastFM} and \textit{MovieLens}. 
Figure \ref{fig:ml_Recommendation} shows the SR@T comparisons on \textit{MovieLens}. We receive the same results in Section \ref{sec:recexp}, where the proposed RN in \modname{} outperforms all other recommendation baselines.

\section{Reproducibility: Experimental Settings}
\label{app:setting}
We first introduce the detailed settings of the hyperparameters.
The initial learning rate is set to 0.001 and we employ Cosine Annealing \cite{Ilya2016} during training, which is a type of learning rate schedule that relatively rapidly decreases the learning rate to a minimum value and then increases it rapidly again. Cosine annealing is empirically proved to have better convergence behavior than linear annealing. The margin $m$ in Eq. \ref{rec_loss} is set to 0.5. We use Adam \cite{kingma2014adam} for training.
Besides, the attribute embedding matrix $E^{\text{attr}}$ is not updated when training the Recommendation Network (RN). We update $E^{\text{attr}}$ using Eq. \ref{attribute_emebdding} every 500 iterations during training.

The dimensions of both user and item embeddings are set to 64. For initializing user and item embeddings, we use \textsc{Leporid} method, where the number of nearest neighbors is set to 1000 and the regularization
coefficient in \textsc{Leporid} is set to 0.5.
To calculate the predictive uncertainty, we perform $N=10$ forward passes. For the threshold $\alpha$, we apply different settings on different datasets: \textit{LastFM}: $\alpha = 0.1$; \textit{MovieLens}: $\alpha = 0.3$; \textit{Yelp}: $\alpha = 0.1$.

\section{Reproducibility: Network Architectures}
\label{app:network}
In this section, we introduce the detailed network architecture of Belief Tracking Network (BTN) and Recommendation Network (RN) in \modname{}.

For BTN, both convolutional layers are set to: kernel size = 3, stride = 1, output channel size = 64. Then we add a fully connected layer where the output dimension is set to 128. After concatenating the output with $\bm{e}^{\text{user}}_u$, the concatenation is fed to three fully connected layers, with the output dimension 512, 1024 and $P^2$, respectively. The final output dimension is strictly set to $P^2$ as to reshape to the matrix $\tilde{A} \in \mathbb{R}^{P\times P}$.

For RN, we stack two residual blocks in the Residual Network. In
the first residual block, we have two convolutional layers (kernel
size = 3, stride = 1, output channel size = 64), activated by ReLU.
In the second block, we employ convolutional layers (kernel size
= 3, stride = 2, output channel size = 128). 
Then we concatenate the output with $\bm{o}_t$ and fed the concatenation into two fully connected layers to generate $\bm{s}_{t}$ , where the output dimension for each layers is deigned to 256, $D$, respectively. 
The item embedding of the item to be evaluated and $\bm{s}_{t}$ are feed to three fully connected layers to predict a score indicating the distance between the user preference and the evaluated item. The output dimension for each fully connected layer is set to 256, 128, 1, respectively.

\section{Reproducibility: Random Seed}
Readers may replicate the results reported in the paper using the random seed of 123. 

\end{document}